\documentclass[twocolumn,showpacs,preprintnumbers,amsmath,amssymb]{revtex4}

\usepackage{graphicx}
\usepackage{dcolumn}
\usepackage{bm}

\begin{document}


\newcommand{\kbar}{$\bar{K}$}
\newcommand{\km}{$K^-$}

\title{The basic $\bar{K}$ nuclear cluster $K^- pp$\\
 and its enhanced formation in the $p + p \rightarrow K^+ + X$ reaction}

\author{Toshimitsu Yamazaki}
\altaffiliation{Department of Physics, University of Tokyo, Bunkyo-ku, Tokyo 113-0033, Japan, and
RIKEN Nishina Center, Wako, Saitama 351-0198, Japan}
\thanks{E-mail address: yamazaki@nucl.phys.s.u-tokyo.ac.jp} 
\author{Yoshinori Akaishi}
\altaffiliation{College of Science and Technology, Nihon University, Funabashi, Chiba 274-8501, Japan, and
RIKEN Nishina Center, Wako, Saitama 351-0198, Japan}
\thanks{E-mail address: akaishi@post.kek.jp}
 
\date{\today}

\begin{center}
Phys. Rev. C, in press\\
\end{center}

\begin{abstract}
 We have studied the structure of $K^- pp$ nuclear cluster comprehensively by solving this three-body system exactly in a variational method starting from the Ansatz that the $\Lambda(1405)$ resonance $(\equiv \Lambda^*)$ is a $K^-p$ bound state. We have found that our original prediction for the presence of $K^-pp$ as a compact bound system with $M = 2322$ MeV/$c^2$, $B_K = 48$ MeV and $\Gamma = 60$ MeV remains unchanged by varying the $\bar{K}N$ and $NN$ interactions widely as far as they reproduce $\Lambda(1405)$. The structure of $K^-pp$ reveals a molecular feature, namely, the $K^-$ in $\Lambda^*$ as an ``atomic center" plays a key role in producing strong covalent bonding with the other proton. We have shown that the elementary process, $p + p \rightarrow K^+ + \Lambda^* + p$, 
which occurs in a short impact parameter and with a large momentum transfer ($Q \sim 1.6$ GeV/$c$), leads to unusually large self-trapping of $\Lambda^*$ by the participating proton, since the $\Lambda^*$-$p$ system exists as a compact doorway state propagating to $K^-pp$ ($R_{\Lambda^* p} \sim 1.67$ fm). 
\end{abstract}


\maketitle

\section{Introduction}

Recently, exotic light nuclear systems involving a $\bar{K}$  ($K^-$ and $\bar{K}^0$) as a constituent have been predicted based on phenomenologically constructed $\bar{K}N$ interactions \cite{Akaishi:02,Yamazaki:02,Dote:04a,Dote:04b,Yamazaki:04,Akaishi:05,Kienle:06}. The predicted bound states in
$K^-ppn$, $K^-ppnn$ and $K^-$$^8$Be with large binding energies lie below the $\Sigma \pi$ emission threshold, and thus are expected to have relatively narrow decay widths. Because of the strong $\bar{K}N$ attraction they acquire enormously high nucleon densities, $\rho_{\rm av} \sim 0.5$ fm$^{-3}$, about 3 times the normal nuclear density $\rho_0  \sim 0.17$ fm$^{-3}$. Such compact nuclear systems, which can be called  ``$\bar{K}$  nuclear clusters" ({\it KNC}), are often those formed with non-existing nuclei. The basic ingredient for this new family of nuclear states is the $I=0~K^-p$ state, which is identified to the known $\Lambda (1405)$ resonance (hereafter, expressed as $\Lambda^*$) in the $\Sigma\pi$ channel with a binding energy of $B_K$ = 27 MeV and a width of $\Gamma$ = 40 MeV \cite{Dalitz}. Since the $\Lambda (1405)$ resonance is largely populated
     in the $p + K^- \rightarrow \Lambda^* + (\pi \pi)^0$ channel \cite{Alston:61}. it is very likely to be the $I=0~\bar{K} N$ state. This is also supported by the large formation of $\Lambda^*$ in the $K^-$ absorption at rest on $^4$He \cite{Riley:75} and also in nuclear emulsion \cite{Davis:77},

The lightest system following this ``$\Lambda(1405)$ Ansatz"  is $K^-pp$ (and its isospin partner $\bar{K}^0 pn$), which was predicted to exist with $M = 2322$ MeV/$c^2$, $B_K$ = 48 MeV and $\Gamma$ = 61 MeV  \cite{Yamazaki:02}. This species, which can be called {\it kaonic dibaryon} or {\it nuclear kaonic hydrogen molecule}, results from a fusion of $\Lambda^*$ and $p$, namely, $\Lambda^*$ as a bound state of $K^- p$ ``dissolves" into a $\bar{K}$  bound state, $K^- pp$, as 
\begin{equation}
 \Lambda^* + p \rightarrow K^-pp, 
\end{equation}
where $\Lambda^*$ may or may not keep its original structure. The situation resembles the diatomic molecule case, where the hydrogen atom (H = $p e^-$) cannot exists as it is when it merges with a proton into a $p$-$e^-$-$p$ (H$_2^+$) molecule. The hydrogen atom, when implanted into a solid, becomes  ``hydrogen in solids", where the hydrogen takes various forms, such as deep/shallow donors and ionized states. It is extremely interesting to ask to what extent the $\Lambda^*$ keeps its identity in nuclear systems. This question is related to the proposal that the $\Lambda^*$ plays a role as a doorway to form $\bar{K}$ bound states \cite{Yamazaki:02}.

In the present paper we first study the $K^- pp$ composite as a very unique three-body system in which an exotic particle ($K^-$) plays an unusually peculiar role in the three-body dynamics through the very strong attractive interaction in $K^- p$. The study was carried out by solving this three-body system exactly in a variational method, called ``Amalgamation of Two-body correlations into Multiple
Scattering process (ATMS)" \cite{Akaishi:86}, which is a method to construct a realistic wave
function of few-body system with correlation functions of each
constituent pairs on the basis of Watson's multiple scattering theory \cite{Watson:53}.
The correlation functions are variationally determined from a given
Hamiltonian by using Euler-Lagrange's equation. We used the elementary $\bar{K}N$ and $NN$ interactions deduced semi-empirically to obtain not only the binding energy and width but also the spatial and momentum distributions of the individual particles. We justify our three-body calculations by showing that the $\bar{K}N$ complex potential, which is transformed from coupled-channel interactions, has very little energy dependence. Furthermore, we show that the result remains unchanged, even when we allow the $\bar{K} N$ and $NN$ interactions to vary in a wide range, as long as they reproduce the energy and width of $\Lambda(1405)$. Thus we are led to a robust consequence that the predicted $K^- pp$ is a compact nuclear system with a binding energy around  50 MeV and a rms $p-p$ distance of 1.9 fm. Through this study we have found that the $K^-p$ unit (quasi-$\Lambda^*$) behaves like an atomic unit in a ``molecule" of $K^- pp$, similar to the mechanism of the Heitler-London scheme \cite{Heitler:27}. Namely, a super strong nuclear force is caused by a migrating real $\bar{K}$ meson, as pointed out in \cite{Yamazaki:07}. This will be the central subject of section \ref{sec:structure}. The decay property of $K^-pp$ has been studied theoretically in  \cite{Ivanov:06}, which showed that, in addition to the dominant decay process to $\Sigma \pi N$, the  partial decay rate to $YN$ is around 20 MeV.
Recently, Faddeev calculations have been carried out to obtain the pole of $K^-pp$ by Shevchenko {\it et al.} \cite{Shevchenko:06} and by Ikeda and Sato \cite{Ikeda:06}. Their pole values are close to our original result. 

Some indications for $K^-pp$ were reported in the invariant-mass spectrum of $\Lambda+p$. An old propane bubble chamber experiment with several GeV proton and neutron beams showed a peak at $M_{\rm inv} (\Lambda p) \sim 2260$ MeV/$c^2$ \cite{Aslanyan}. A more recent experiment of FINUDA at DAPHNE on stopped-$K^-$ reactions on light nuclei revealed a peak at $M_{\rm inv} (\Lambda p) = 2250$ MeV/$c^2$ \cite{FINUDA:PRL}. This result was interpreted by the experimental group as indicating a bound $K^-pp$ state with $B_K \sim 110$ MeV, whereas two different  theoretical arguments have been published \cite{MORT,Yamazaki:07b}. This issue has to wait for further confirmation by future experiments. It is important to produce various $\bar{K}$ clusters by different nuclear reactions and thereby to examine their structure, formation and decay properties. A method to determine the sizes of the $\bar{K}$  clusters via the momentum correlation of decay particles has been proposed \cite{Kienle:06}. 

In the second part of the present paper we study the possibility to make use of  the elementary process,
\begin{equation}
p + p \rightarrow p + \Lambda^* + K^+,
\end{equation}\label{eq:pp}
in which $\Lambda^*$ and $p$ proceed to $K^- pp$. Since the momentum transfer in this associated production of $\Lambda^*$ is very large ($Q \sim$ 1.6 GeV/$c$), one would expect that the formation cross section of $K^- pp$ must be very small. This process resembles the hypernuclear production process, $^A [Z](p, K^+)^{A+1}_{\Lambda}[Z]$, on a nuclear target, the cross section of which was evaluated by Shinmura {\it et al.}  \cite{Shinmura} to be $10^{-4}$ of the elementary $\Lambda$ production cross section, even when a short-range correlation is taken into account. On the other hand,  with a naive coalescence mechanism one obtains a sticking probability of the order of 0.1-1.0 \% because the internal momentum of the $\bar{K}$ clusters is very large \cite{Suzuki-Fabietti}. Still, most of primarily produced $\Lambda^*$ are expected to escape, and the quasi-free process dominates. We have studied this proton-induced associated production process more realistically, and found a surprisingly large production cross section by a unique mechanism, as described in section \ref{sec:pp-reaction}. Its preliminary description is seen in \cite{Yamazaki:06a} in connection with an experiment proposal at GSI using the FOPI detector \cite{FOPI-proposal}. Short communications of the present results are also seen in Ref. \cite{Yamazaki:06b,Yamazaki:07}

\section{Structure of $K^- pp$}\label{sec:structure}

\subsection{The bare $\bar{K}N$ interactions}

We start from the Ansatz that the $\Lambda(1405)$ resonance state is the $I=0$ 1s bound state of $\bar{K} N$. Through the main part of this paper we employ the ``classical" experimental values for the binding energy and width \cite{Dalitz,Alston:61},
\begin{eqnarray}
-B_K &=& E_{\bar{K}N}^{I=0} =  -27~ {\rm MeV},\\
\Gamma &=& 40~{\rm MeV}.
\end{eqnarray}\label{eq:Lambda*}
Later, in subsection \ref{subsec:beyond}, we will make a fine tuning, considering recent values, $M = 1406 \pm 4$ MeV and $\Gamma = 50.0 \pm 2.0$ MeV \cite{PDG}.

The $\Lambda^*$ data, \ref{eq:Lambda*}, combined with the kaonic hydrogen shift \cite{Iwasaki:97,Beer:05} (yielding $a_{K^- p}$) and Martin's $\bar{K} N$ scattering lengths ($a^{I=0}$ and $a^{I=1}$) \cite{Martin},  
\begin{eqnarray}
a_{K^- p}&=&(-0.78 \pm 0.15) + i\, (0.49 \pm 0.28)~{\rm fm}, \\
a^{I=0}&=&(-1.70 \pm 0.07) + i\,(0.68 \pm 0.04)~{\rm fm},\\
a^{I=1}&=&(0.37 \pm 0.09) + i\, (0.60 \pm 0.07)~{\rm fm},
\end{eqnarray} 
were used in a coupled-channel calculation to deduce the $\bar{K} N$ interactions of the following forms \cite{Akaishi:02}
\begin{eqnarray}
v_{\bar{K} N} ^I &=& v_D \, {\rm exp}[-(r/b)^2], \\
v_{\bar{K} N, \pi \Sigma } ^I &=& v_{C_1} \, {\rm exp}[-(r/b)^2], \\
v_{\bar{K} N, \pi \Lambda} ^I &=& v_{C_2} \, {\rm exp}[-(r/b)^2], 
\end{eqnarray} 
where 
\begin{equation}
b = 0.66~{\rm fm}
\end{equation}
and $v_D^{I=0} = -436$ MeV, $v_{C_1}^{I=0} = -412$ MeV, $v_{C_2}^{I=0}$ = none, $v_D^{I=1} = -62$  MeV, $v_{C_1}^{I=1} = -285$ MeV, $v_{C_2}^{I=1} = -285$ MeV. The two interactions, $v_{\pi \Sigma}^{I} (r)$ and $v_{\pi \Lambda}^{I} (r)$, are taken to be vanishing to simply reduce the number of parameters. 
This is justified because they are almost irrelevant in describing the \kbar~bound states.

\begin{figure}
\centering
\includegraphics[width=\columnwidth]{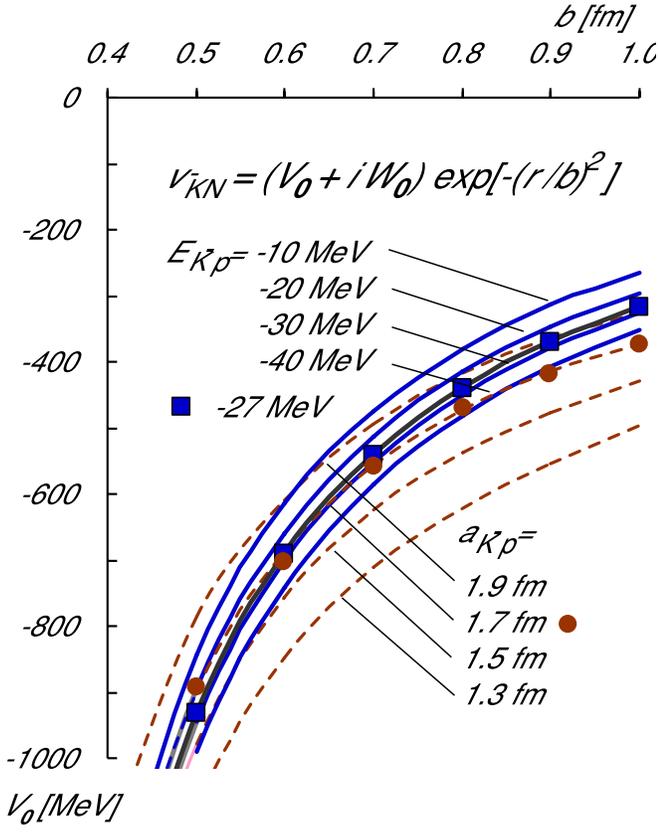}
\vspace{0cm}
\caption{\label{fig:V-b-contour} 
(Color online) Parametric presentation of the $K^- p$ energy, $E_{K^- p}$, and the scattering length, $a_{K^- p}$, in the plane of the $\bar{K} N$ interaction strength ($V_0$) and the range parameter ($b$) in the expression $v^{I=0}_{\bar{K} N}(r) = (V_0 + i\, W_0)\,{\rm exp} [-(r/b)^2]$. The imaginary part is adjusted so as to reproduce $\Gamma = 40$ MeV. The experimental values, $E_{K^- p}$ = -27 MeV (blue squares) and 
$a_{K^- p}$ = 1.7 fm (red circles), determine $V_0$ and $b$. 
}
\end{figure}

The above coupled-channel interactions were used to derive equivalent
single-channel $\bar{K} N$ potentials with imaginary parts in
energy-independent forms, which is an appropriate way to obtain the {\it 
decaying state} of Kapur-Peierls
\cite{Kapur:38} as discussed below. The obtained complex potentials are:
\begin{eqnarray}
v_{\bar{K}N}^{I=0} (r) &=& (-595 - { i} \, 83)\, {\rm exp} [-(r/0.66)^2], \label{eq:vKN0} \\
v_{\bar{K}N}^{I=1} (r) &=& (-175 - { i} \, 105)\, {\rm exp} [-(r/0.66)^2],\label{eq:vKN1}
\end{eqnarray}
in units of MeV and fm. The same range is assumed for $I=0$ and $I=1$.
The interaction strength ($V_0$) and the range ($b$) can be determined simultaneously because $B$ and $a_{K^- p}$ have different dependences on $V_0$ and $b$, as shown in Fig.~\ref{fig:V-b-contour} (and also in Table \ref{tab:KN-range}). Our semi-empirical $\bar{K} N$ interaction is consistent with the theoretically derived ones from meson-exchange \cite{Mueller:90} and from chiral dynamics \cite{Kaiser:97,Cieply:01,Borasoy:05}.

\begin{figure}
\centering
\includegraphics[width=\columnwidth]{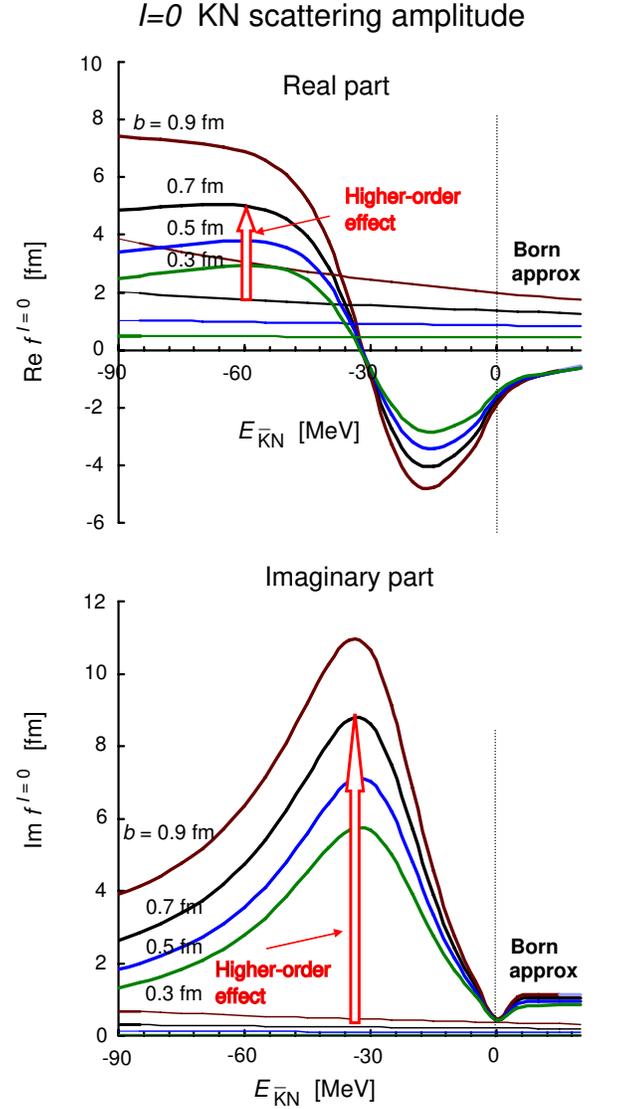}
\vspace{0cm}
\caption{\label{fig:scatt-amp} 
(Color online) Dependence of the $\bar{K} N$ scattering amplitude on the range parameter $b$. (Upper) The real part with $b = 0.7$ fm reproduces the known chiral dynamics result \cite{Kaiser:97}. (Lower) The imaginary part with $b = 0.7$ fm accounts for the observed total cross section of $\Lambda (1405)$. 
}
\end{figure}

Noting that the parameter $b$ in the above Gaussian distribution is related to the rms distance $R$ as $b = \sqrt{2/3} R = 0.816 R$, we find the observed proton rms radius ($R_p = 0.862$ fm) to give a range parameter $b = 0.70$ fm, which is compatible with our range parameter (0.66 fm). To see further consistency we have calculated the $\bar{K} N$ scattering amplitude by changing the range parameter $b$. The results are shown in Fig.~\ref{fig:scatt-amp}. The real and imaginary parts with $b = 0.7$ fm reproduces the chiral dynamics result \cite{Kaiser:97,Cieply:01,Borasoy:05} very well, in
spite of the strong claim by Oset and Toki \cite{Oset:06}  that AY's scattering
amplitudes are too large compared with those obtained from the chiral
unitary approach of Oset and Ramos \cite{Oset-Ramos:98}.
 Thus, the interaction range deduced and used in AY is fully justified.

Now, Let us discuss the energy dependence of the single-channel complex $\bar{K} N$ potential. We employ Yukawa-type separable potentials as the original coupled-channel interaction to treat the problem analytically, which are
\begin{eqnarray}
\langle \vec k' \mid v_{ij} \mid \vec k \rangle = g(\vec k') U_{ij}^{(0)} g(\vec k), ~~~g(\vec k) = \frac {\Lambda^2} {\Lambda^2 + \vec k^2}, \\
U_{ij}^{(0)} = \frac{1}{\pi^2} \frac{\hbar^2}{2 \sqrt{\mu_i \mu_j}}
\frac{1}{\Lambda} s_{ij}.
\end{eqnarray} 
where $i,j$ stand for the $\bar{K} N$ channel (1) or the $\pi \Sigma$ channel (2), and 
$s_{ij}$ are non-dimensional strength parameters. The experimental binding energy and width of $\Lambda(1405)$ are reproduced with $s_{11}=-1.022,~s_{12}=-0.626,~s_{22}=0$ and $\Lambda = 770$ MeV$/\hbar c = 3.9$ fm$^{-1}$. This range corresponds to a Gaussian range, $b = 2/\Lambda = 0.51$ fm, which is consistent with the Gaussian range $b=0.66$ fm we use.

\begin{figure}
\centering
\includegraphics[width=\columnwidth]{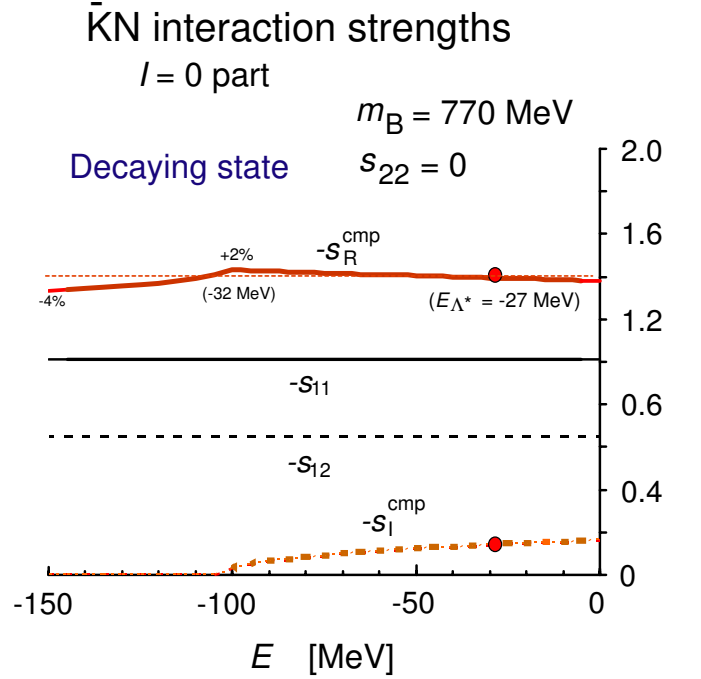}
\vspace{0cm}
\caption{\label{fig:KN-energy-dep} 
(Color online) Energy dependence of the $\bar{K} N$ interaction strength, where $s^{\rm cmp} = s_{\rm R}^{\rm cmp} + i\, s_{\rm I}^{\rm cmp}$.
}
\end{figure}

The single-channel complex $\bar K N$ potential can be derived by Feshbach's projection operator procedure:
\begin{eqnarray}
v^{\rm {cmp}}(E) &=& Pv_{11}P \nonumber \\
 &+&Pv_{12}Q \frac {1}{E-Qh_{22}Q+i\epsilon} Qv_{21}P,
\end{eqnarray}
namely,
\begin{eqnarray}
s^{\rm {cmp}}(E) &=& s_{11}-s_{12} \frac {\Lambda^2}{(\Lambda-i\kappa_2)^2+s_{22}\Lambda^2} s_{21},\\
~~E+\Delta M &=&\frac {\hbar^2}{2 \mu_2^2} \kappa_2^2,
\end{eqnarray} 
where $P$ and $Q$ are the projection operators to the $\bar K N$ channel and the $\pi \Sigma$ channel, respectively, and $\Delta M$ is the threshold mass difference between the two channels. In case of non-zero $s_{22}$ a virtual state sometimes appears on the $\pi \Sigma$ unphysical sheet, which gives a serious energy dependence of $v^{\rm {cmp}}(E)$, when used to obtain the "pole state" of $K^-pp$ in three-body Faddeev calculations \cite{Shevchenko:06,Ikeda:06}. However, one should notice that experimental observation is done {\it not for the "pole state" but for the "decaying state"}, as understood from the open-channel asymptotic behavior of Green's function of Morimatsu-Yazaki \cite{Morimatsu-Yazaki} describing the process of $K^-pp$ production reactions. The energy dependence of the single-channel $\bar K N$ potential (real part) is only a little for the "decaying state" as shown in Fig.~\ref{fig:KN-energy-dep}. The imaginary part describing the decaying state decreases to zero toward the $\Sigma \pi$ threshold, as physically expected, whereas we used the energy independent potentials in the calculation of $K^- pp$.  This decrease of the imaginary part changes the width of $K^-pp$ from 61 to 43 MeV, whereas the larger width of $\Lambda(1405)$ from 40 to 50 MeV causes a canceling effect; from 43 to 54 MeV, as shown later in subsection \ref{subsec:beyond}.

Thus, our energy-independent potentials of Eq.(\ref{eq:vKN0},\ref{eq:vKN1}) are justified with sufficient accuracy, demonstrating that it is {\it just a proper way} of treating the experimentally observable {\it decaying state} of $K^-pp$. 
Nevertheless, we will examine in the next section (\ref{sec:validity-test}) how the $K^-pp$ structure depends on different choices of the $\bar{K} N$ and $NN$ interactions by varying the interaction parameters widely.

\subsection{The bound state of $K^- pp$}

The presence of a deeply bound dibaryonic $\bar{K}$ system, $K^- pp$, was first predicted as a natural extension of $K^- p$ in \cite{Yamazaki:02}. A variational method (ATMS) developed in \cite{Akaishi:86} was employed together with the bare $\bar{K} N$ interaction of AY \cite{Akaishi:02} and the bare $NN$ interaction of Tamagaki \cite{Tamagaki},
\begin{eqnarray}
&&v_{NN} (r) = 2000 \,{\rm exp}[-(r/0.447)^2] \nonumber \\
&&- 270 \, {\rm exp}[-(r/0.942)^2] -5 \, {\rm exp}[-(r/2.5)^2].
\end{eqnarray}
In these expressions we have employed the length units in fm and the energy units in MeV.

The three-body variational wave function of $\bar{K}NN$ with a number definition $(1, 2, 3) = (\bar{K}, N, N)$ is given as 
\begin{equation}
\Psi = [\Phi_{12} + \Phi_{13}]\,|T=1/2>
\end{equation}
where
\begin{eqnarray} \label{eq:Phi}
\Phi_{12} &=& [f^{I=0} (r_{12})\, P_{12}^{I=0}  + f^{I=1} (r_{12})\, P_{12}^{I=1}]\nonumber \\
&\times& f_{NN} (r_{23}) f(r_{31}),\\
\Phi_{13} &=& f (r_{12}) f_{NN}(r_{23})  \nonumber \\
&\times& [f^{I=0} (r_{31})\, P_{31}^{I=0}  + f^{I=1} (r_{31})\, P_{31}^{I=1}], 
\end{eqnarray}
with
\begin{eqnarray}
&&P_{12}^{I=0} = \frac{1 - \vec{\tau_K} \cdot \vec{\tau_N}}{4},\\
&& P_{12}^{I=1} = \frac{3 + \vec{\tau_K} \cdot \vec{\tau_N}}{4}.
\end{eqnarray}
The functions $f^{I=0} (r_{ij})$ and $f^{I=1} (r_{ij})$ are scattering correlation functions of the particle pair $(i,j)$ for the $I=0$ and $I=1$ $\bar{K}N$ interactions, respectively, and $f_{NN} (r_{23})$ is that for the $NN$ pair, and $f(r_{i,j})$ is for the off-shell case.  
The $T=1/2$ state consists of three isospin eigenstates as
\begin{eqnarray}
&&|T=1/2> =  \sqrt{\frac{3}{4}} \, \biggl[(\bar{K}_1 N_2)^{0,0} \, p_3 \biggr] \\ \nonumber
&&+ \sqrt{\frac{1}{4}} \, \biggl[ - \sqrt{\frac{1}{3}} (\bar{K}_1 N_2)^{1,0} \, p_3 
+ \sqrt{\frac{2}{3}} (\bar{K}_1 N_2)^{1,1} \, n_3 \biggr], 
\end{eqnarray}
where $(\bar{K}_1 N_2)^{I, I_z}$ is for the isospin $(I, I_z)$. Among these the first term corresponds to $\Lambda^* p$.

\begin{figure}
\centering
\includegraphics[width=0.95\columnwidth]{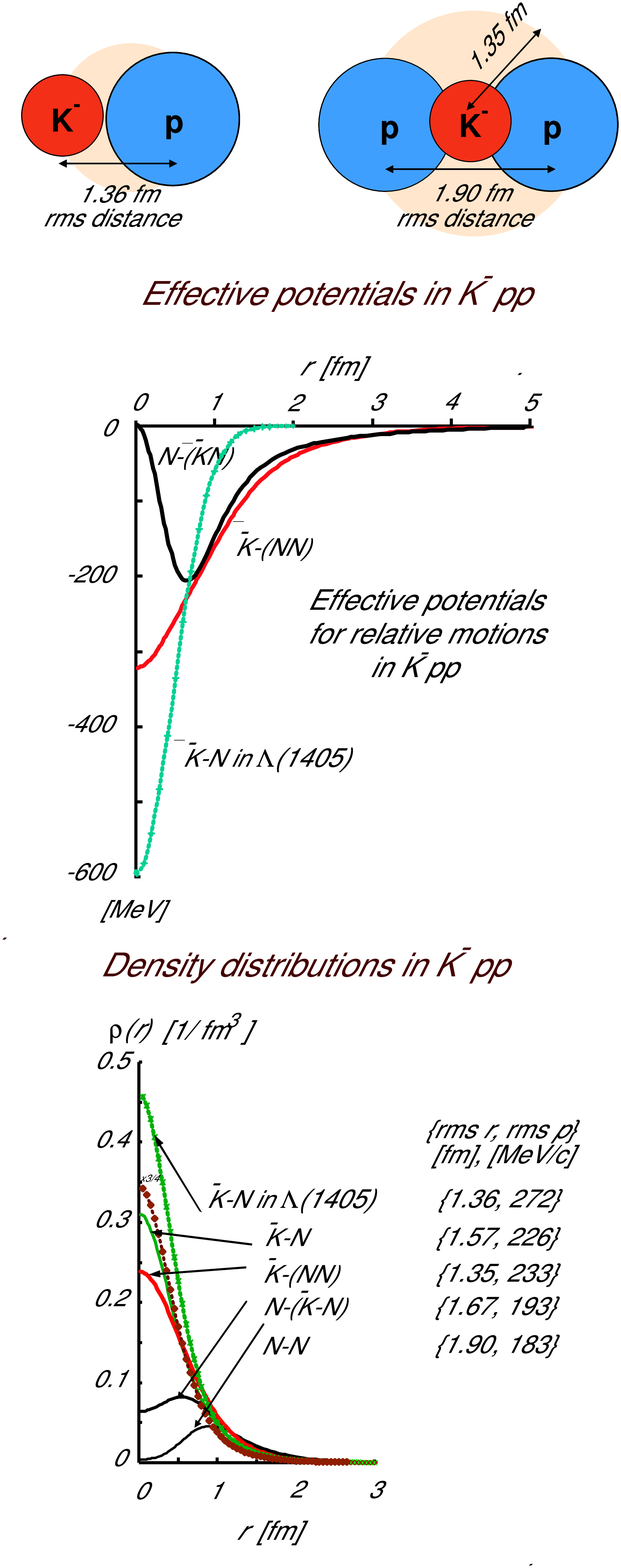}
\vspace{-0.5cm}
\caption{\label{fig:Structure-Kpp} 
(Color online)
(Upper) Structure of $K^-p$ and $K^-pp$, as calculated in \cite{Yamazaki:02}. 
(Middle) The effective potentials for relative motions of $N$-$(\bar{K}N)$ and $\bar{K}$-$(NN)$, deduced from the exact variational wavefunction for $K^-pp$. The $K^-$-$p$ potential for $\Lambda(1405)$ is also shown. (Lower) Density distributions of various coordinates in $K^-pp$ as well as in $\Lambda(1405) = K^-p$ together with its density reduced by a factor of 3/4 (brown dots). The values of the rms distances and momenta are also given.}
\end{figure}

The binding energy and width of $K^-pp$ thus obtained are:
\begin{eqnarray}
-B_K &=& E_{\bar{K}NN} =  -48~ {\rm MeV},\\
\Gamma &=& 61~{\rm MeV}.
\end{eqnarray}
For the estimate of the width we have taken into account only the pionic decay modes of $\bar{K}N \rightarrow Y \pi$. The width will be larger if we consider other decay modes such as $K^-pp \rightarrow YN$, which have been studied theoretically \cite{Ivanov:06}.

The predicted structure of $K^-p$ and $K^-pp$ is shown in Fig.~\ref{fig:Structure-Kpp}. The wavefunction of $\Lambda^* = K^-p$ in our treatment is expressed by $\phi_{\Lambda^*} (r)$ with $r$ being the $K^-$-$p$ distance. Its density distribution is shown in Fig.~\ref{fig:Structure-Kpp}. The rms distance of $K^-$-$p$ is 1.36 fm.
The ``nucleus" $pp$ does not exist, but the $K^-$ can combine two protons into a strongly bound system, when they are in a spin-singlet state. The predicted state is expressed as $K^- (pp)_{S=0, T=1}$, and its isospin partner is $K^- (pn)_{S=0, T=1}$, or more generally, $[\bar{K} (NN)_{S=0,T=1}]_{T=1/2}$. It was shown in \cite{Akaishi:02} that the normal deuteron ($S=1,T=0$) does not form a deeply bound state with $K^-$,  but a non-existing ``excited deuteron" of $I=1$ can do. These results come from the three-body variational calculation, but can easily be understood in terms of the different weights of the $I=0$ and $I=1$ $\bar{K}N$ interactions in the di-baryonic configurations:
\begin{eqnarray}
~[K^- \times (nn)_{S=0}]_{T=3/2}&:&~~2\,[v^{I=1}],\\
~[K^- \times (d)_{S=1}]_{T=1/2}&:&~~2\,[\frac{1}{4}v^{I=0} + \frac{3}{4} v^{I=1}],\\
~[K^- \times (pp)_{S=0}]_{T=1/2}&:&~~2\,[\frac{3}{4} v^{I=0} + \frac{1}{4} v^{I=1}].
\end{eqnarray}  
Namely, the third one, $K^-pp$ (and its isobaric analog state), has the deepest energy level.

\subsection{Density distributions in $K^-pp$}

Here we show and discuss the calculated density distributions of $K^- pp$ in details. The effective potential energies as functions of the relative distances of $\bar{K}$-$(NN)$ and $N$-$(\bar{K} N)$ are extracted from the obtained total wave function, as shown in Fig.~\ref{fig:Structure-Kpp} (Middle). 
The distributions of the relative distances and the momenta of the constituent particle pairs, namely, 
$\bar{K}$-$N$, $\bar{K}$-$(NN)$, $(\bar{K} N)$-$N$, and $N$-$N$, were calculated. Figure \ref{fig:Structure-Kpp} (Lower) shows their density distributions, $\rho (r)$. The calculated rms distances and rms momenta are also presented. The $N$-$N$ rms distance is 1.90 fm, which is significantly smaller than the average inter-nucleon distance in normal nuclei, and is much smaller than the rms distance of $p$-$n$ in $d$ (3.90 fm). The $N$-$(\bar{K} N)$ potential has a core followed by a strong attactive part, and the $N$-$(\bar{K} N)$ distribution yields a rms distance of 1.67 fm. The rms radius of $\bar{K}$ with respect to $(NN)$ is 1.35 fm, close to the rms distance of $\bar{K}$-$N$ in $\Lambda(1405)$.

\begin{figure}
\centering
\includegraphics[width=\columnwidth]{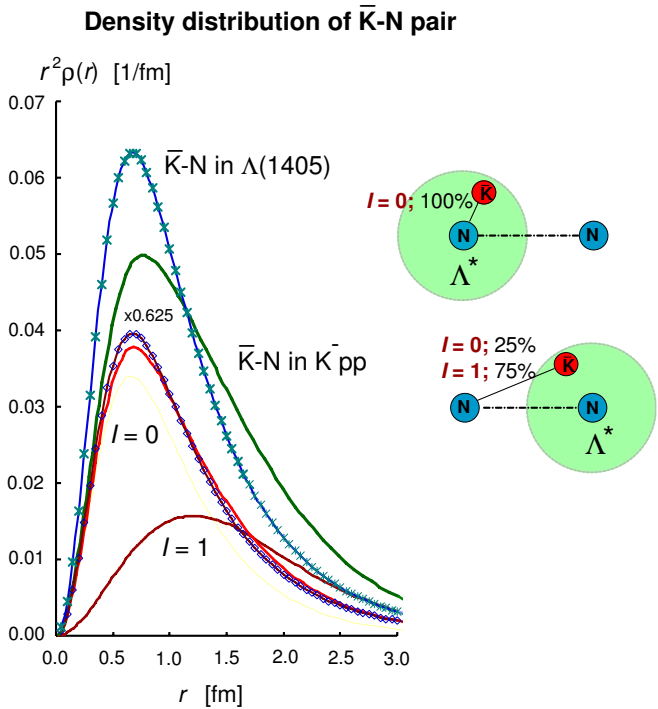}
\vspace{0cm}
\caption{\label{fig:KbarN-density} 
(Color online) Comparison of the density distributions, $r^2\,\rho(r_{KN})$, of the $\bar{K}$-$N$ distance in the  $\bar{K} N$ pair in $\Lambda (1405)$ and in $K^-pp$. The latter is decomposed into the $I=0$ and $I=1$ pairs. The density distribution in $\Lambda(1405)$ after multiplication of a factor 0.625 is also shown. }
\end{figure}

It is interesting to see how the original structure of the $K^-$-$p$ binding in $\Lambda(1405)$ persists in $K^-pp$. For this purpose, we compare in Fig.~\ref{fig:Structure-Kpp} and Fig.~\ref{fig:KbarN-density} the $\bar{K}$-$N$ distance distributions of the $\bar{K} N$ pair in $K^-pp$, $\rho_{\bar{K}-N}(K^-pp)$, with that in $\Lambda(1405)$, $\rho_{\bar{K}-N}(\Lambda^*)$. Most naively, we would expect that $\rho_{\bar{K}-N}(K^-pp) = (3/4) \rho_{\bar{K}-N}(\Lambda^*)$, as shown by brown dots in Fig.~\ref{fig:Structure-Kpp} (Lower), if $K^-$ were bound by one of the two protons, resulting in a free $\Lambda(1405)$ and a proton, whereas the realistic calculation indicates that the former ($R^{\rm rms}_ {\bar{K}-N}$ = 1.57 fm) is significantly broader than the latter (1.36 fm). This can be qualitatively understood, since the original $\bar{K} N$ pair is dissolved into the three-body system of $K^-pp$. To investigate this difference more deeply, we decompose the density distribution into the $\bar{K} N^{I=0}$ and $\bar{K} N^{I=1}$ parts, as shown in Fig.~\ref{fig:KbarN-density}. The $I=0$-pair distribution has a shape closer to $\rho_{\bar{K}-N}(\Lambda^*)$, whereas the $I=1$ part is widely distributed due to the smaller attractive interaction.  

Although the shape of $\rho_{\bar{K}-N}(K^-pp)$ is similar to that of $\rho_{\bar{K}-N}(\Lambda^*)$, their intensities are different. This can be understood as follows. When $K^-$ (1) resides with Proton (2) with a probability of 0.5, the $I=0$ component of the wave function $\Phi_{12}$ in (\ref{eq:Phi}) dynamically increases to 1 due to the strong $\bar{K} N^{I=0}$ interaction. We also expect an additional intensity ($0.5 \times 1/4 = 0.125$) from Proton (3),  and the total intensity becomes $0.625 \, \rho_{\bar{K}-N}(\Lambda^*)$, which accounts for $\rho_{\bar{K}-N}(K^-pp)$ very well. This means that $K^-$ (1) in $K^-pp$ resides partially around Proton (2) in a form of $\Lambda (1405)$, and partially around Proton (3), as given by the total wave function. This indicates that the structure of $\Lambda(1405)$ is nearly unchanged when it dissolves into this ``nucleus". In other words, the $\Lambda(1405)$ state, though modified, persists in a nuclear system. This aspect  justifies the $\Lambda(1405)$ doorway model \cite{Yamazaki:02}. 

In analyses of the J\"{u}lich group \cite{Mueller:90}, 
the $\bar {K} N$ interaction in the relevant energy region was found to be mainly 
of the $t$-channel type, where $\omega$, $\rho$ and $\sigma$ meson exchanges 
coherently contribute to the strong $I=0$ attraction which is enough to 
accommodate a bound state assigned to $\Lambda(1405)$. 
By taking into account the dominance of this interaction 
we describe
the $K^- pp$ and $K^- pn$ subsystems as $\Lambda^* p$ and $\Lambda^* n$, and thus the decay interaction as $\Lambda^* p \rightarrow \Lambda p$ ({\it ``proton participant"} case) and $\Lambda^* n \rightarrow \Lambda n$ ({\it ``neutron participant"} case), respectively.

\subsection{Molecular aspect of $K^-pp$}

The persistency of $\Lambda^*$ in $K^-pp$ reminds us of a molecular type binding, similar to hydrogen molecule, $p + e^- + p$  (H$_{2}^+$) and muonic hydrogen molecule, $p + \mu^- + p$ ($\mu^-$H$_{2}^+$). The inter-atomic distance scale is given by the Bohr radius, $a_{\rm B} = 0.53$ nm, and the scale of the muonic molecule is by $a_{\mu} = 256$ fm. The present kaonic nuclear cluster $K^-pp$ can be interpreted as a kaonic hydrogen molecule in the sense that $K^-$ traverses between the two protons, producing ``strong covalency" through the strongly attractive $\bar{K}N^{I=0}$ interaction. This is essentially the mechanism of Heitler and London \cite{Heitler:27} for hydrogen molecule, though the nature of the interaction is totally different and the mass of the migrating particle is much heavier.

\begin{figure}[]
\centering
\includegraphics[width=\columnwidth]{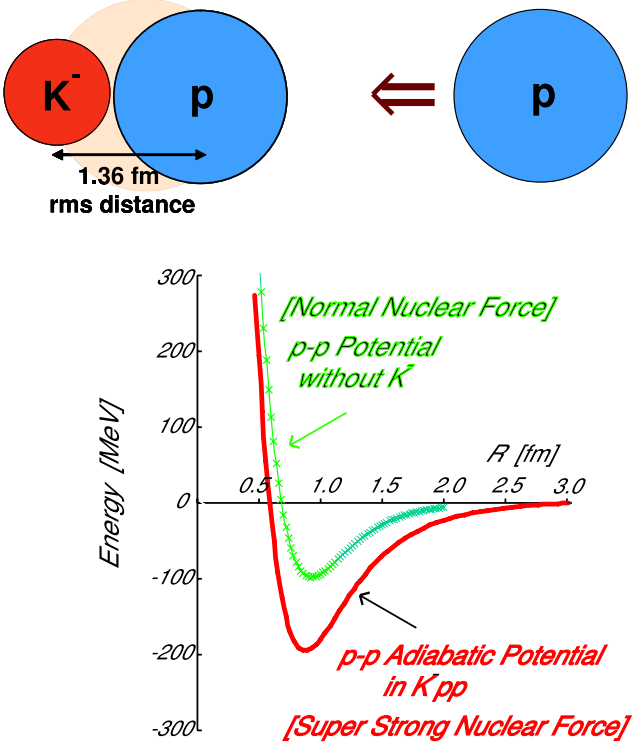}
\vspace{0cm}
\caption{\label{fig:AdiaPot} 
(Color online) The adiabatic potential when a proton approaches a $\Lambda(1405)$ as a function of the $p$-$p$ distance. For comparison the Tamagaki potential $v_{NN}$ \cite{Tamagaki} is shown.}
\end{figure}

This aspect is more clearly seen, when the density distribution is plotted with a fixed axis of the two protons. Figure~\ref{fig:AdiaPot} shows the adiabatic potential, when a proton approaches a $\Lambda(1405)$ particle, as a function of the $p$-$p$ distance. The $p$-$p$ potential caused by the migrating $K^-$ is much deeper than the bare $p$-$p$ interaction. This can be called ``super strong nuclear force", as compared with the ordinary nuclear force. When a $\Lambda^*$ is produced in a close proximity with a proton, it easily binds the proton. This leads to a $\Lambda^* p$ doorway situation following the $\Lambda^*$ doorway, as will be discussed later.

\begin{figure}
\centering
\includegraphics[width=\columnwidth]{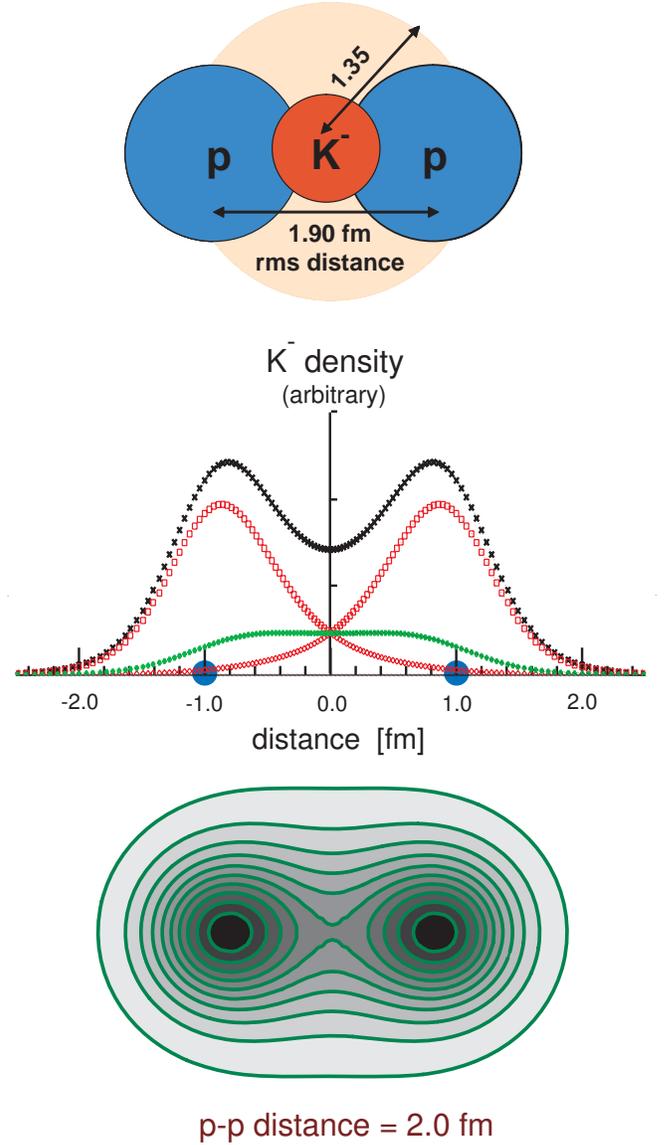}
\vspace{0cm}
\caption{\label{fig:molecule} 
(Color online) The molecular structure of $K^-pp$. (Middle) The projected density distributions of $K^-$ in $K^-pp$ with a fixed $p$-$p$ distance (= 2.0 fm). (Lower) The corresponding $K^-$ contour distribution.}
\end{figure}

Figure~\ref{fig:molecule} shows the projected distribution of $K^-$ along the $p$-$p$ axis and the contour distribition of $K^-$, when the $p$-$p$ distance is fixed to 2.0 fm. This case resembles the ground state of $K^-pp$, as the calculated rms distance is 1.9 fm. From these figures we recognize the distinct character of $K^-pp$ as a ``diatomic molecule". Namely, the $K^-$ is distributed not around the center of $p$-$p$, but around each of the two protons. The $K^-$ distribution is composed of the ``atomic" part, as shown by curves of red open circle chain, and the exchange part by green broken curve.

It is interesting to see how the individual energy terms behave in the light of the Heitler-London picture. The ``atomic" system, $K^-p$, has $E = -27.8 - {\rm i}\,20$ MeV, $<T_K> = 115.3$ MeV, and $<v_{\bar{K} N}> = 143.1 -{\rm i}\, 20.0$ MeV. The $K^-$-$(pp)$ part in the molecular system, $K^-pp$, has $<T_K> = 118.3$ MeV, $<2 v_{\bar{K}N}> = -195.5$ MeV and thus, $E = -77.2 - {\rm i}\, 30.6$ MeV. On the other hand, the $p$-$p$ interaction part has $<T_{NN}> = 48.8$ MeV, $<v_{NN}> = -19.0$ MeV, and $E = 29.8$ MeV. Thus, the energy  difference attained when the molecular state is formed from the atomic state is as follows:
\begin{eqnarray}
\Delta E = -47.5 + 27.8 = -19.7~{\rm MeV},\\
\Delta T_{\bar{K}} = 118.3 - 115.3 = 3.0~{\rm MeV},\\
\Delta E_{NN} = 48.8 - 19.0 = 29.8~{\rm MeV},\\
\Delta V_{\bar{K}N} = -195.5 + 143.1 = -52.4~{\rm MeV}. 
\end{eqnarray}
Since the massive $K^-$ causes a shrinkage of $pp$, the $pp$ energy increases together with the $K^-$ kinetic energy. Nevertheless, the strong $I=0$ $\bar{K}N$ attraction produces a large exchange integral,
\begin{eqnarray}
 \sum _{\{i,j\} = \{2,3\},\{3,2\}}  && <\Phi_{1i} |v_{\bar{K}N} (12) + v_{\bar{K}N} (13) |\Phi_{1j}>  \nonumber
\\  
&&= -52.6~{\rm MeV},
\end{eqnarray}
which is the source for the deeper binding of $K^-pp$ as compared with $K^-p$.

\subsection{Super strong nuclear force caused by a migrating  real $K^-$}

Despite the drastic dynamical change of the system caused by the strong $\bar{K} N$ interaction the identity of the ``constituent atom", $\Lambda^*$, is nearly preserved because of the presence of a short-range repulsion between the two protons. This extremely dense ``molecule" can be called ``sub-femto mini-molecule". In the same sense, the previously predicted $K^-K^-pp$ \cite{Yamazaki:04} corresponds to the two-electron neutral hydrogen molecule (H$_2^0$). 

Historically, Heisenberg \cite{Heisenberg:32} tried to explain the origin of the strong nuclear force in terms of ``Platzwechsel", namely, $n \leftrightarrow p + e^-$, as in the molecular bonding, originating from Heitler and London \cite{Heitler:27}, but had to abandon this idea for obvious reasons. Then, Yukawa introduced a mediating virtual meson \cite{Yukawa:35}. This hypothetical meson was later discovered, and Yukawa's idea of ``mediating boson" was established as the fundamental concept  in the contemporary particle physics, the most notable being the $W$ and $Z$ weak bosons. It is to be noted that the $K^-pp$ system (and subsequent kaonic clusters) is regarded as a revival of the Heitler-London-Heisenberg scheme, where a super strong nuclear force is produced by a migrating real boson, $K^-$, as emphasized in \cite{Yamazaki:07}, where the volume integral of the super strong nuclear force is by a factor 4.1 larger than that of the ordinary nuclear force.

\section{Validity test against various potential parameters}\label{sec:validity-test}

One may raise a question: how robust are these predictions on $K^- pp$?  
In the following we study comprehensively the effect of the $\bar{K} N$ and $NN$ interactions by varying them to wide extent, while reproducing the energy and width of $\Lambda(1405)$. 

\subsection{Dependence on the $NN$ hard core}

\begin{figure}
\centering
\includegraphics[width=\columnwidth]{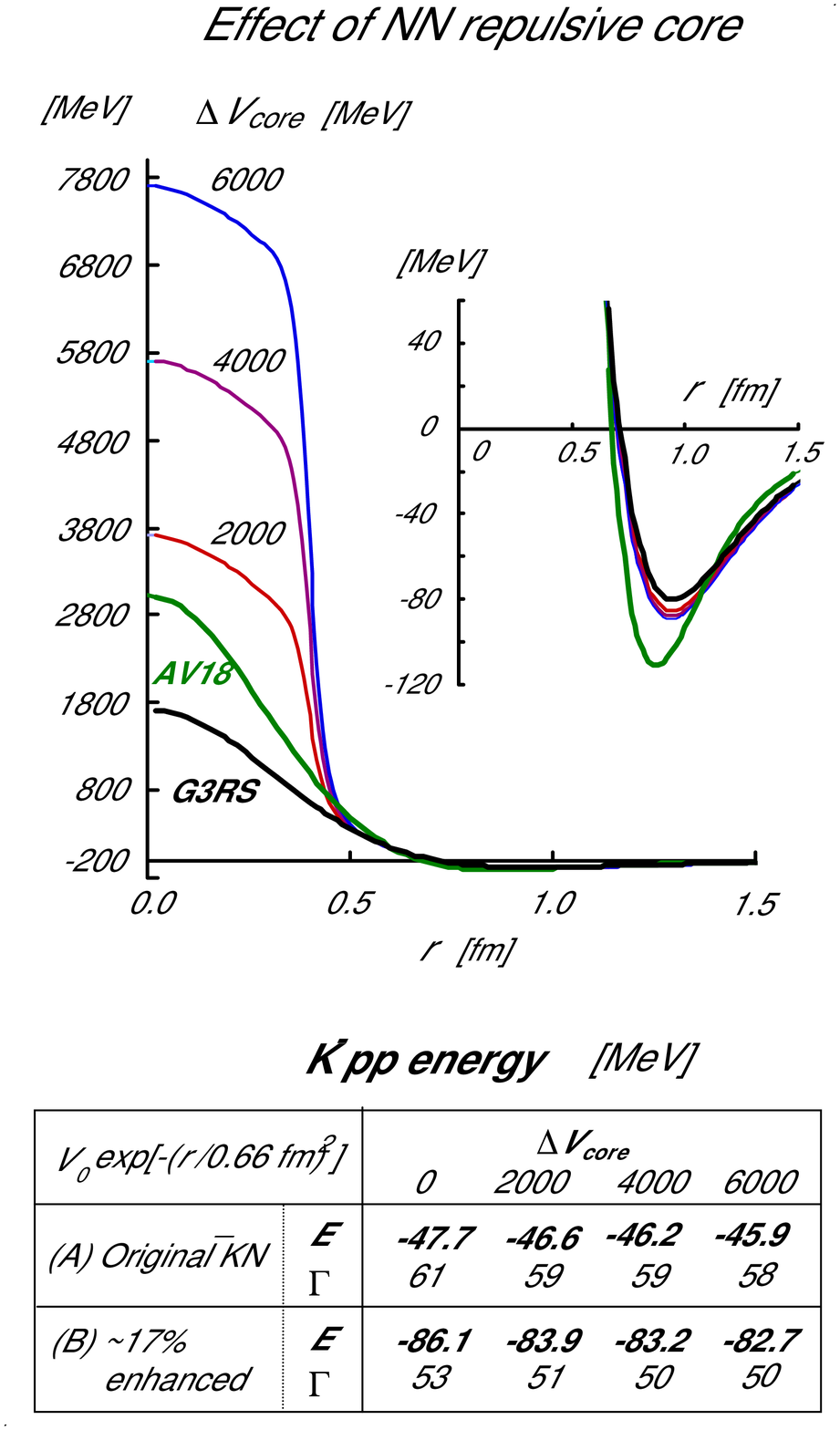}
\vspace{0cm}
\caption{\label{fig:pot_hardcore} 
(Color online) The artificially increased hard core in the $NN$ potential, while keeping the $NN$ scattering lengths to empirical values. The calculated energy and width of $K^-pp$ with varied hard core values $V_{\rm core}$ are listed in inset. 
}
\end{figure}

First, one may wonder how the $NN$ hard core will affect the binding of $K^- pp$. To examine this effect we introduce an unrealistically large hard core by adding 
\begin{equation}
\frac{\Delta V_{\rm core}}{1 + (r/0.4~{\rm fm})^{20}}
\end{equation}   
with $\Delta V_{\rm core}$ = 2000, 4000, and 6000 MeV to the original Tamagaki potential G3RS \cite{Tamagaki}, 
\begin{eqnarray}
&&v_{NN}^{\rm G3RS} = 2000\,{\rm exp}[-(r/0.447)^2] \nonumber\\
 && + v_2\,{\rm exp}[-(r/0.942)^2] - 5 \,{\rm exp}[-(r/2.5)^2]
\end{eqnarray}
with $r$ in fm, as shown in Fig.~\ref{fig:pot_hardcore}. For reference the Argonne potential AV18 \cite{Wiringa:95} is also shown. To assure self consistency the mid-range attractive strength ($v_2$), the original value of which is $-270$ MeV, was adjusted so as to reproduce the $NN$ scattering length. Namely, $v_2 =$ -270, -284, -291 and -295  MeV for $\Delta V_{\rm core}$ = 0, 2000, 4000, and 6000 MeV, respectively. Then, the binding energy and width turned out to change only slightly: $B_K$ = 47.7, 46.6, 46.2 and 45.9 MeV for respective $\Delta V_{\rm core}$.

\subsection{Dependence on the $K^- N$ interaction range}

\begin{figure}
\centering
\includegraphics[width=7cm]{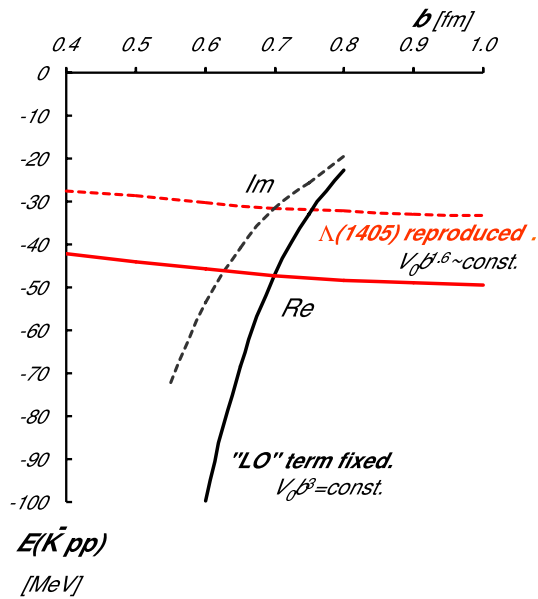}
\vspace{0cm}
\caption{\label{fig:b-dependence} 
(Color online) Dependence of the $K^-pp$ energy and width on the range parameter $b$ in the present treatment (red curves), where the interaction strength is varied so as to reproduce the energy and width of $\Lambda(1405)$. For comparison is shown the case of  the chiral dynamics treatment with a fixed ``LO" term (black curves). 
}
\end{figure}

In order to examine the effect of the interaction range let us vary the range ($b$) of the $\bar{K} N$ interaction in a form of
\begin{equation}
v_{\bar{K} N}^{I=0} = (V_0 + i\, W_0)\,{\rm exp}[-(\frac{r}{b})^2]
\end{equation}
drastically from 0.3 to 1.0 fm while reproducing the $K^- p$ binding to the observed $B_K$ and $\Gamma$ of $\Lambda(1405)$. The results are shown in Table~\ref{tab:KN-range}. Obviously, $|V_0|$ (and also $|W_0|$) increases with the decrease of $b$. On the other hand, the energy and width of $K^-pp$ do not change much. This situation is shown in Fig.~\ref{fig:b-dependence}. Table~\ref{tab:KN-range} also shows a gradual change of the $I=0$ scattering length with $b$, which clarifies and confirms our statement in connection with Fig.~\ref{fig:V-b-contour}
 that $B$ and $a_{K^- p}$ have different dependences on $V_0$ and $b$.

In a chiral dynamics derivation of the $\bar{K}N$ interaction \cite{Kaiser:97} the interaction range in the expression of the form factor as
\begin{equation}
v(k;b) = C_{\rm ``LO"} \, {\rm exp}[-b^2k^2/4]
\end{equation}
is chosen so as to be consistent with the energy and width of $\Lambda(1405)$. This means that the constant $C_{\rm ``LO"}$ is determined according to the chosen $b$. When the range parameter $b$ is varied, the $C_{\rm ``LO"}$ parameter is kept constant in usual treatments. Then, the derived potential strength is varied so as to fulfill the relation: $V_0 b^3$ = const. Such a treatment does, however, not reproduce the $\Lambda(1405)$ as $K^- p$ and yield a totally different $b$ dependence in the energy and width of $K^-pp$, as shown in the same figure. To reproduce the $\Lambda(1405)$ energy the interaction strength should fulfill a different relation, $V_0 b^{1.6} \approx$ const.

\begin{table}[h]
\caption{\label{tab:KN-range}  Calculated potential parameters ($V_0$ and $W_0$ in MeV), energies ($E_{K^- pp}$) and widths ($\Gamma_{K^- pp}$) of $K^-pp$ in MeV, and the $I=0$ scattering length in fm with varied $\bar{K} N$ range ($b$ in fm), while reproducing $\Lambda(1405)$.  }
\begin{center}
\begin{tabular}{cccccc}
\hline
$b$  &  $V_0$ & $W_0$ & $E_{K^- pp}$   & $\Gamma_{K^- pp}$   & $a^{I=0}$ (fm)  \\
\hline
1.0   & $-316.5$         & $-62.0$ &  $-49.5$         &  66.5    &  $-1.95 + i\, 0.45$ \\
0.9   & $-368.7$         & $-67.0$  & $-49.0$         &  65.7     & $-1.89 + i\, 0.44$\\
0.8   & $-439.6$         & $-73.0$  & $-48.3$         &  64.4    & $-1.82 + i\, 0.44$ \\
0.7   & $-540.0$         & $-81.0$  & $-47.3$         &  62.9   & $-1.75 + i\, 0.43$  \\
0.6   & $-689.5$         & $-91.0$  & $-45.8$         &  60.3    & $-1.69 + i\, 0.43$ \\
0.5   & $-929.7$         & $-105.0$  & $-44.0$         &  57.4   & $-1.62  + i\, 0.42$ \\
0.4   & $-1358.0$         & $-128.0$  & $-42.1$         &  54.9  & $-1.55 + i\, 0.42$ \\  
0.3  & $-2250.0$        & $-162.0$  & $-40.1$       & 51.3      & $-1.48 + i\, 0.42$\\
\hline
\end{tabular}
\end{center}
\end{table}

In order to examine whether or not the $E_{K^- pp}$ and $\Gamma_{K^- pp}$ depend on the functional form of the interaction we take the Yukawa type form:
\begin{equation}
v_{\bar{K} N} ^{I=0} = (V_0 + i\, W_0) \, \frac{b_Y}{r}\,{\rm exp}[-\frac{r}{b_Y}],
\end{equation}
where the range parameter $b_Y$ is related to the Gaussian parameter $b$ as
\begin{equation}
b_Y = 0.5 \, b.
\end{equation}
For each value of $b_Y$ the parameters $V_0$ and $W_0$ were determined so as to reproduce $\Lambda(1405)$. The results are shown in Table~\ref{tab:KN-range-Yukawa}. Here again, the $E_{K^-pp}$ and $\Gamma_{K^- pp}$ do not change much. They are found to be close to those obtained with the Gaussian type interaction.

\begin{table}[h]
\caption{\label{tab:KN-range-Yukawa}  Calculated potential parameters ($V_0$ and $W_0$ in MeV) and energies ($E_{K^- pp}$) and widths ($\Gamma_{K^- pp}$) of $K^-pp$ in MeV with varied $\bar{K} N$ Yukawa range ($b_Y$ in fm), while reproducing $\Lambda(1405)$.  }
\begin{center}
\begin{tabular}{ccccc}
\hline
$b_Y$  &  $V_0$ & $W_0$ & $E_{K^- pp}$   & $\Gamma_{K^- pp}$     \\
\hline
0.50   & $-593.3$         & $-60.7$ &  $-45.1$         &  57.0     \\
0.45   & $-709.8$         & $-67.7$  & $-44.4$         &  56.2     \\
0.40   & $-869.5$         & $-76.6$  & $-43.7$         &  55.3     \\
0.35   & $-1098.0$         & $-88.0$  & $-43.0$         &  54.3     \\
0.30   & $-1444.0$         & $-103.2$  & $-42.4$         &  53.4     \\
0.25   & $-2006.0$         & $-124.8$  & $-41.9$         &  52.5     \\
0.20   & $-3021.0$         & $-157.5$  & $-41.8$         &  52.0   \\  
\hline
\end{tabular}
\end{center}
\end{table}

\begin{figure}
\centering
\includegraphics[width=7cm]{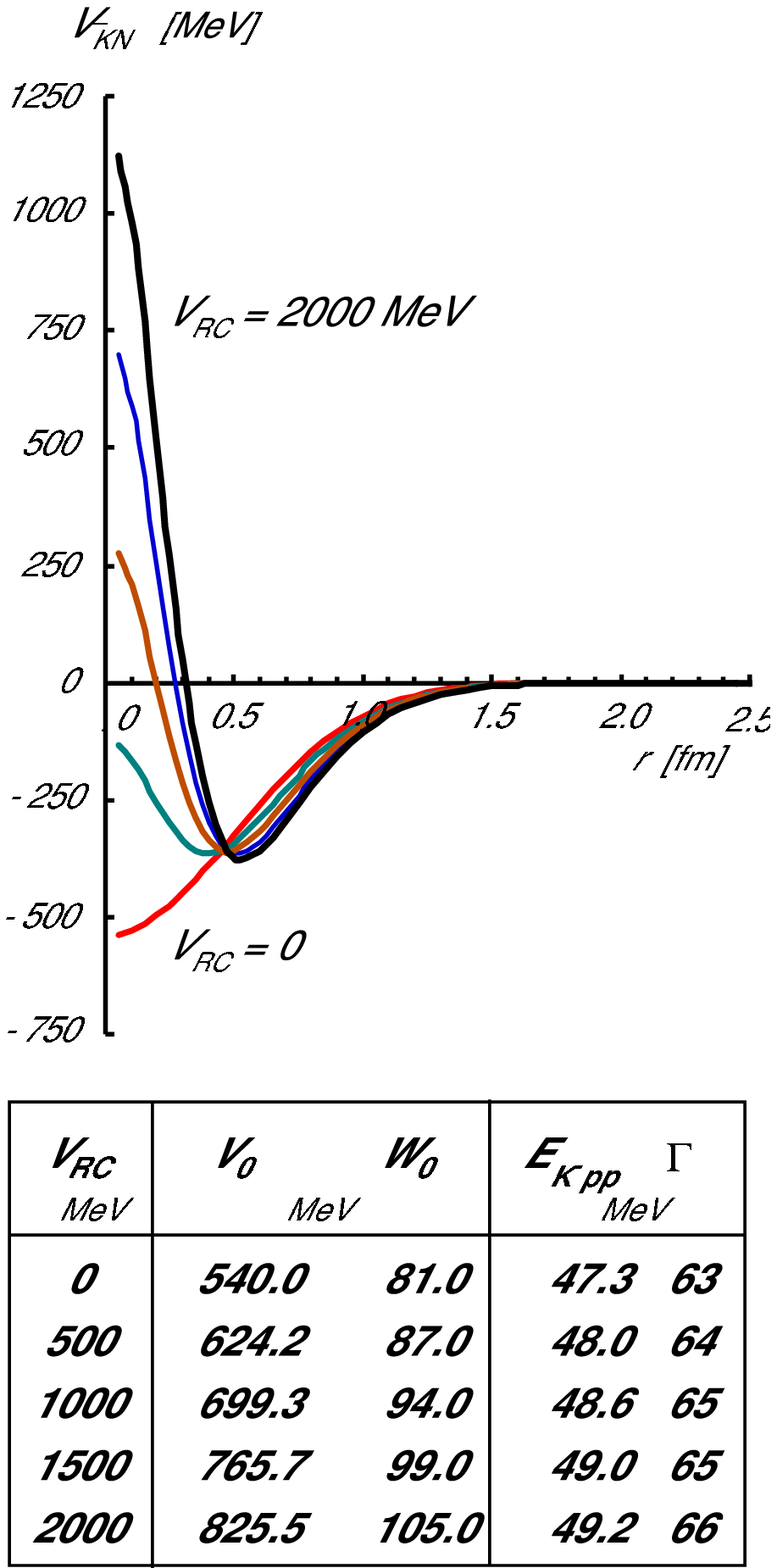}
\vspace{0cm}
\caption{\label{fig:KN_hardcore} 
(Color online) The $\bar{K} N$ potentals with artificially introduced hard core values, $V_{\rm RC}$ = 0, 500, 1000, 1500 and 2000 MeV, while keeping the energy and width of $K^- p$ to the $\Lambda^*$ values. The calculated potential parameters and energy and width of $K^-pp$ with varied hard core values $V_{\rm RC}$ are listed in inset. }
\end{figure}

\subsection{Dependence on the $K^- N$ hard core}

No hard core has been imposed so far for the $\bar{K} N$ interaction. Here, we attempt to invoke the following $\bar{K} N$ interaction with a hard-core part to examine its effect on $K^- pp$ binding:
\begin{eqnarray}
v_{\bar{K} N} &=& V_{\rm core}\,{\rm exp}[-(\frac{r}{b_1})^2] \nonumber \\
&& + (V_0 + i\, W_0) \,{\rm exp}[-(\frac{r}{b_2})^2]  
\end{eqnarray}
with $b_1$ = 0.3 fm and $b_2$ = 0.7 fm. The $\bar{K} N$ potential was changed with various $V_{\rm core}$ values from 0 to 6000 so as to reproduce $\Lambda(1405)$, as shown in Fig.~\ref{fig:KN_hardcore}. The calculated potential parameters ($V_0$ and $W_0$) and the energy and width of $K^-pp$ are listed in its inset. This result also indicates that the energy and width depend only slightly on the assumed hard core in the $\bar{K} N$ interaction. This behavior is unchanged for a different value of $b_2$.

\subsection{Beyond the original prediction of $K^-pp$}\label{subsec:beyond}

 
We have shown in Fig. \ref{fig:KN-energy-dep} that the imaginary part of the $\bar{K} N$ interaction for the decaying state decreases to zero as the energy approaches the $\Sigma \pi$ emission threshold, whereas our original value (61 MeV) for $\Gamma$ of $K^- pp$ does not include this effect. Now, after taking into account the energy dependence of the imaginary part, we obtain a corrected value, $\Gamma \approx  43$ MeV. If we adopt the PDG value (50 MeV) for the width of $\Lambda(1405)$ instead of the earlier value of 40 MeV, we end up with a value, $\Gamma \approx 54$ MeV. This is close to our original value of 61 MeV.
 
Thus, we have confirmed that the original prediction for  $K^- pp$ is quite robust against any change of the two-body interactions involved, as far as the interactions are constrained by $\Lambda (1405)$. If our prediction turns out to be different from future observation, it will indicate anything beyond the present model. In fact, the recent FINUDA experiment \cite{FINUDA:PRL} suggested the presence of  a bound state deeper than the original prediction, and we should keep our eyes open to unknown effects which may come into the three-body system. As such we list the following.\\

i) {\bf Validity of the $\Lambda(1405)$ Ansatz}. Since our treatment depends entirely on this Ansatz, the result would be substantially changed, if this Ansatz were not right (even partially). Non-observation of $K^-pp$ would cast a serious question on the so far believed nature of $\Lambda(1405)$, requiring a totally new physics regarding this resonance.\\

ii) {\bf Presence of a p-wave $\bar{K} N$ interaction}, which is not relevant to $\Lambda(1405)$, but may be  pertinent to three-body (or more) systems \cite{Wycech}. \\

iii) {\bf The $\bar{K} N $ interaction modified in three-body (or more) systems}, such as due to chiral symmetry restoration or other QCD effects.\\

iv) {\bf The $NN$ repulsion relaxed by the presence of $\bar{K}$.}  If the short-range $NN$ repulsion results from the $uud$-$uud$ interaction, the intruding $K^- = s \bar{u}$ brings a $\bar{u}$ in between, and the $N$-$N$ repulsion may be weakened by a kind of shielding, namely, $uud$-$s \bar{u}$-$uud$. \\

In view of these unknown effects we have to be prepared to predict the effect of widely varied $\bar{K} N$ interaction (without constraint by $\Lambda(1405)$). Specifically, we consider the following three different cases for the $\bar{K} N$ interaction: \\

\noindent
(A) the original AY interaction, \\
(B) enhanced interaction strength by 1.17, and \\
(C) enhanced interaction strength by 1.25. \\

The cases (B) and (C) were adopted, when we discussed the possible change of the $\bar{K} N$ interaction corresponding to different observations of the tribaryonic $\bar{K}$ bound states \cite{Akaishi:05}. The bound-state energies and widths, and mutual rms distances, momenta and densities of the three bodies, $\bar{K}$, $N$ and $N$, in $K^-pp$ are presented in Table~\ref{tab:KNN}. In the next sections we employ the cases (A) and (B) for production reactions.

\begin{table}
\caption{\label{tab:KNN}  Calculated energies and widths in units of MeV of $K^-pp$ with three different $\bar{K} N$ interactions. $<KE>$: average kinetic energy. $<PE>$: average potential energy. The rms distances, momenta and densities among the three bodies are also shown. }
\vspace{0.5cm}

(A) Original AY\\
~~~~~$E = -48 - {\rm i}\, 30$, ~ $<KE> = 162, ~ <PE> =-210$
\begin{tabular}{lccc} 
\hline
                                     &  $\bar{K}$-$(NN)$ & $N$-$(\bar{K}N)$   & $N$-$N$     \\
\hline
rms $R$ [fm]              & 1.35          & 1.67              &   1.90     \\  
rms $P$ [MeV/$c$]   & 233          & 193             &  183  \\
$\rho (0)$  [fm$^{-3}$] & 0.24      & 0.062          & 0.007 \\
\hline
\end{tabular}
\vspace{0.5cm}

(B) 17\% enhanced\\
~~~~~$E = -86 - {\rm i}\, 27$, ~ $<KE> = 208, ~ <PE> =-294$
\begin{tabular}{lccc} 
\hline
                                       &  $\bar{K}$-$(NN)$ & $N$-$(\bar{K}N)$   & $N$-$N$     \\
\hline
rms $R$ [fm]              & 1.14          & 1.44              &   1.65     \\  
rms $P$ [MeV/$c$]   & 270          & 218             &  205  \\
$\rho (0)$  [fm$^{-3}$] & 0.35      & 0.079          & 0.012 \\
\hline
\end{tabular}
\vspace{0.5cm}

(C) 25\% enhanced\\
$E = -106 - {\rm i}\, 29$, ~ $<KE> = 228, ~ <PE> =-333$
\begin{tabular}{lccc} 
\hline
                                       &  $\bar{K}$-$(NN)$ & $N$-$(\bar{K}N)$   & $N$-$N$    \\
\hline
rms $R$ [fm]              & 1.08          & 1.37              &   1.58     \\  
rms $P$ [MeV/$c$]   & 285          & 228             &  214  \\
$\rho (0)$  [fm$^{-3}$] & 0.42      & 0.086          & 0.014 \\
\hline
\end{tabular}
\end{table}

\section{Ordinary $\bar{K}$ transfer reactions - $\Lambda^*$ doorway}\label{sec:ordinary-reaction}

The conventional methods to produce kaonic bound states are to use strangeness-transfer reactions of $(K^-,\pi^-)$, $(\pi^+,K^+)$, $(K^-,N)$ and $(\gamma,K^+)$. We treat the formation of $\bar{K}$ clusters by a $\Lambda^*$ doorway model \cite{Yamazaki:02}, in which a $\Lambda^*$ produced in elementary processes, typically,  
\begin{eqnarray}
K^- + n \rightarrow \Lambda^* + \pi^-, \label{eq:Kpi}\\
\pi^+ + n \rightarrow \Lambda^* + K^+,\label{eq:piK}
\end{eqnarray}
merges with a surrounding nucleon (or nucleus) to become a $\bar{K}$ state. 

\begin{figure*}
\centering
\includegraphics[width=16cm]{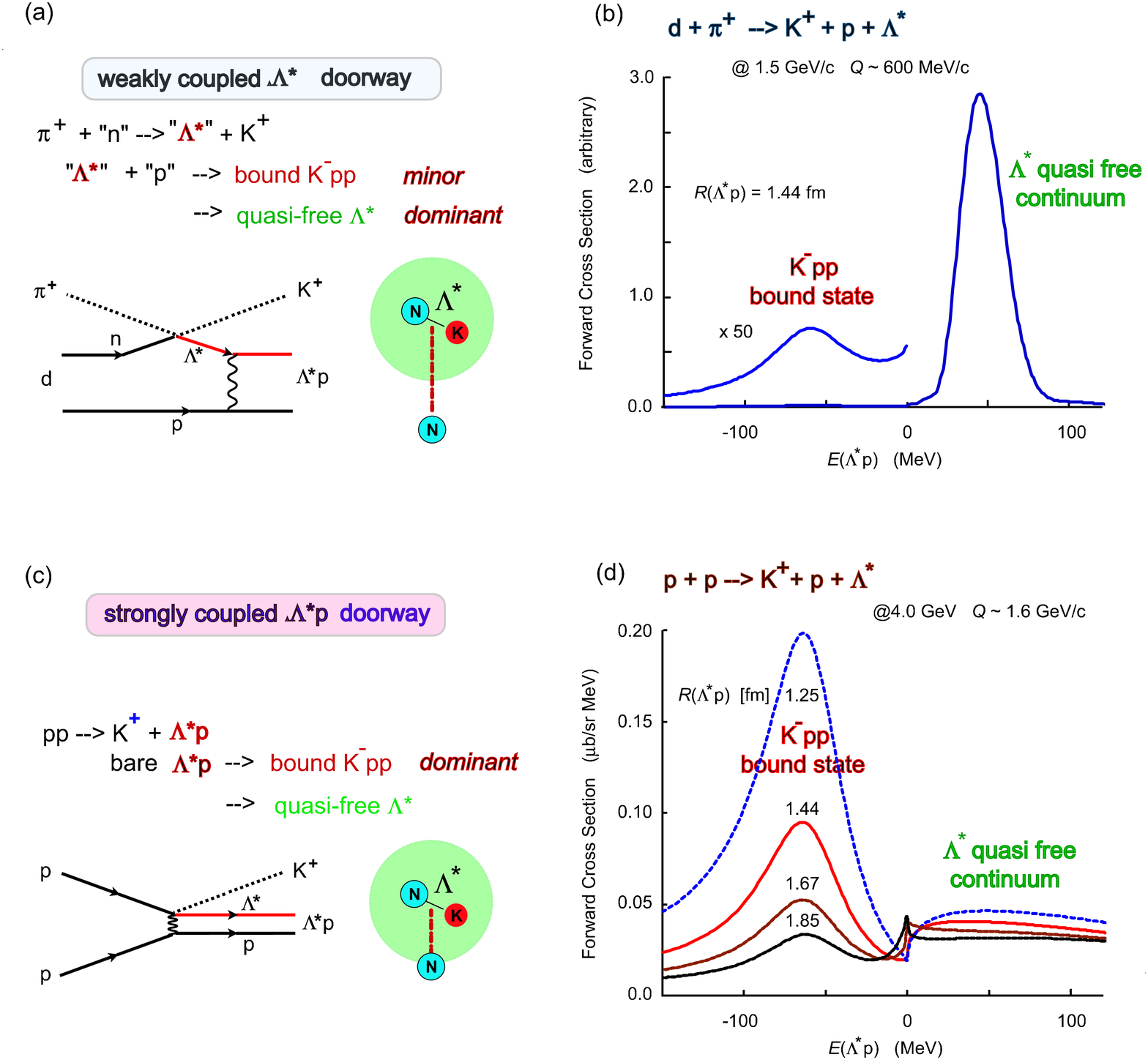}
\vspace{0cm}
\caption{\label{fig:piK-pp-combined} 
(Color online) (a) Diagram for the $d(\pi^+,K^+)K^-pp$ reaction. (b) Calculated spectral shape of the $d(\pi^+,K^+)K^-pp$ reaction. (c) Diagram for the $p(p,K^+)K^-pp$ reaction. (d) Forward cross sections of the $p(p,K^+)K^-pp$ reaction for different rms distances $R(\Lambda^* p)$. The cases (A) and (B) of the $\bar{K} N$ interaction correspond to 1.67 and 1.44 fm, respectively.
}
\end{figure*}

We describe the case of $d(\pi^+,K^+)K^-pp$ reaction. In this case, we use the case (B) with revised  binding energy and width for $K^-pp$ ($B_K = 86$ MeV and $\Gamma = 58$ MeV). In the elementary process, Eq.(\ref{eq:piK}), the produced $\Lambda^*$ interacts with a proton in the target $d$, proceeding to $K^-pp$, as shown in Fig.~\ref{fig:piK-pp-combined} (Upper-Left). The momentum transfer at a typical incident momentum of $p_{\pi} \sim 1.5$ GeV/$c$ is $Q \sim$ 600 MeV/$c$. The energy spectrum involving both the bound and unbound regions was calculated following the Morimatsu-Yazaki procedure \cite{Morimatsu-Yazaki}. It is  given by 
\begin{eqnarray}
  \frac{{\rm d}^2\sigma}{{\rm d}E_{K^+} {\rm d} \Omega_{K^*}} &=& \alpha (k_{K^+}) \frac{{\rm d}\sigma ^{\rm elem}_{\Lambda^*}}{ {\rm d} \Omega_{K^+}} \nonumber \\
&\times&  \frac{| \langle \phi_{\Lambda^*}| v_{\bar{K}N}^{I=0}|\phi_{\Lambda^*} \rangle |^2}{\tilde{E}^2 + \frac{1}{4} \Gamma_{\Lambda^*}^2} S(E)
\end{eqnarray}\label{eq:piK-cross-section}
with a spectral function
\begin{eqnarray}
S(E)&=&(-\frac{1}{\pi}) {\rm Im} \big[\int {\rm d}\vec{r}_{K} {\rm d}\vec{r}'_{K} \tilde{f}^*(\vec{r}_{K}) \nonumber \\
&\times& \langle \vec{r}_{K} |\frac{1}{E-H_{{K}^-{pp}} + i \epsilon}|\vec{r}'_{K} \rangle \tilde{f}(\vec{r}'_{K}) \big],
\label{eq:spectral}
\end{eqnarray} 
where
$\tilde{E}$ is the energy transfer to the $\Lambda^*$-$p$ relative motion in doorway states, 
 and   
$E$ the energy transfer to the $K^-$-$pp$ relative (internal) motion in the $K^-pp$ system, and $\alpha (k_{K^+})$ is a kinematical factor. The function  $\tilde{f}(r)$ is
\begin{equation}
 \tilde{f}(\vec{r}) = 2^3 {\rm e}^{i 2 \beta \vec{q}\vec{r}} C(r) \Phi_{pp}^* (2r) \Psi_{d} (2r)/|\phi_{\Lambda^*} (0)|,
\end{equation}
with 
$\vec{q} = \vec{k}_{\pi^+} -\vec{k}_{K^+}$,   
$\beta = M_{p}/(M_{\Lambda^*} + M_{p})$ and $C(r) = 1 - {\rm exp}[-(r/1.2~{\rm fm})^2]$, and $\Phi_{pp}$ is the $p-p$ relative wave function in $K^- pp$. In this derivation we have used a zero-range approximation for $v_{\bar{K} N} ^{I=0}$ and closure approximation to doorway states.

The calculated spectral function is shown in Fig.~\ref{fig:piK-pp-combined} (Upper-Right). The dominant part is the quasi-free component, in which the produced $\Lambda^*$ escapes, and only a small fraction constitutes a bound-state peak. The peak intensity of the bound $K^-pp$ state depends on the size of the $\Lambda^* p$ system, but not so drastically.

\section{$K^-pp$ production in $NN$ collisions}\label{sec:pp-reaction}

\subsection{Formulation}

Now we consider the following $\Lambda^* p$ doorway process with a projectile proton and a target proton, 
\begin{equation}
p + p  \rightarrow K^+ + (\Lambda^* p), \nonumber
\end{equation} 
which proceeds to the bound-state formation with two-body final states as well as to free $\Lambda^*$ emission (called ``quasi-free" process):
\begin{eqnarray}
         &\rightarrow&    K^+ + K^-pp,\\
         &\rightarrow&    K^+ + \Lambda^* + p  \label{eq:pp2KLp}.
\end{eqnarray}
The formed $K^-pp$ decays not only via the major channel
\begin{equation}
K^-pp \rightarrow \Sigma + \pi + p,
\end{equation}
but also in non-pionic decay channels: 
\begin{eqnarray}
K^-pp &\rightarrow& \Lambda + p \rightarrow p + \pi^- + p, \label{eq:pp2Kpp-decay-1}\\
K^-pp &\rightarrow& \Sigma^0 + p \rightarrow p + \pi^- + \gamma + p, \label{eq:pp2Kpp-decay-2}\\
K^-pp &\rightarrow& \Sigma^+ + n \rightarrow n + \pi^+  + n. \label{eq:pp2Kpp-decay-3}
\end{eqnarray}
The reaction diagram is shown in Fig.~\ref{fig:piK-pp-combined} (Lower-Left). When the incident proton interacts with a neutron in a deuteron target, an analogous process takes place, namely, $p + n \rightarrow K^+ + \Lambda^*  + n$ with an iso-doublet partner $\Lambda^* n$ (=$K^- (pn)_{I=1,S=0}$)  as well as $p + n \rightarrow K^0 + \Lambda^* + p$.  Hereafter, we take the $p + p$ case without loss of generality. 

The $p \rightarrow p + K^- + K^+$ process, where a $K^+K^-$ pair is assumed to be created at zero range from a proton, is of highly off-energy shell ($\Delta E \sim 2 m_K c^2$). This process is realized only with a large momentum transfer to the second proton, which occurs efficiently by a short-range $pp$ interaction. When it is expressed by a Yukawa type interaction, exp$(-r/m_B)/r$ with $m_B$ being the intermediate boson mass, the effective interaction for the elementary process is written as 
\begin{eqnarray}
&& \langle\vec{r}_{K^+(K^- pp')}, \vec{r}_{(K^- p)p'}, \vec{r}_{K^-p}| t | \vec{r}_{pp'} \rangle \nonumber\\
&&= T_0 \int {\rm d} \vec{r} \, F(\vec{r}) \delta (\vec{r}_{K^+(K^- pp')} - \eta \vec{r}) \nonumber\\
&& \times~ \delta(\vec{r}_{(K^- p)p'} - \vec{r}) 
 \delta(\vec{r}_{K^- p}) \delta(\vec{r}_{p p'} - \vec{r})),
\end{eqnarray}
where
\begin{equation}
F(\vec{r}) = \frac{\beta}{r} \, {\rm exp}(-\frac{r}{\beta})
\end{equation}
with $\beta = \hbar/(m_B c) $ and $\eta = M_p/M_{K^-pp}$. 
The $\Lambda^*$ is treated as a quasi-bound state of $K^- p$, and the interaction matrix element for $\Lambda^*$ formation is given by 
\begin{eqnarray}
&& \langle \vec{k}_{K^+ (\Lambda^* p')} , \vec{r} = \vec{r}_{\Lambda^* p'}, \phi_{\Lambda^*} | t | \vec{k}_{p p'} \rangle \nonumber\\
&&
= T_0 \, \phi_{\Lambda^*} (0)\, F(r)  \langle \vec{k}_{K^+(\Lambda^* p')} | \eta \vec{r} \rangle \langle\vec{r} | \vec{k}_{pp'}\rangle \nonumber\\
&& \equiv U_0 f (\vec{r}),
\end{eqnarray}
where 
\begin{eqnarray}
U_0 &=& \frac{1}{(2 \pi)^3} T_0 \phi_{\Lambda^*} (0),\\
f(\vec{r}) &=& \frac{\beta}{r} {\rm exp} (-\frac{r}{\beta} + {i} \vec{Q} \vec{r}),\\
\vec{Q} &=& \eta_0 \vec{k}_p - \eta \vec{k}_{K^+}, \\
\eta_0 &=& \frac{1}{2} + \frac{m_K}{M_{K^-pp} + m_K} \eta.
\end{eqnarray}

\begin{figure*}
\centering
\vspace{0cm}
\includegraphics[width=15cm]{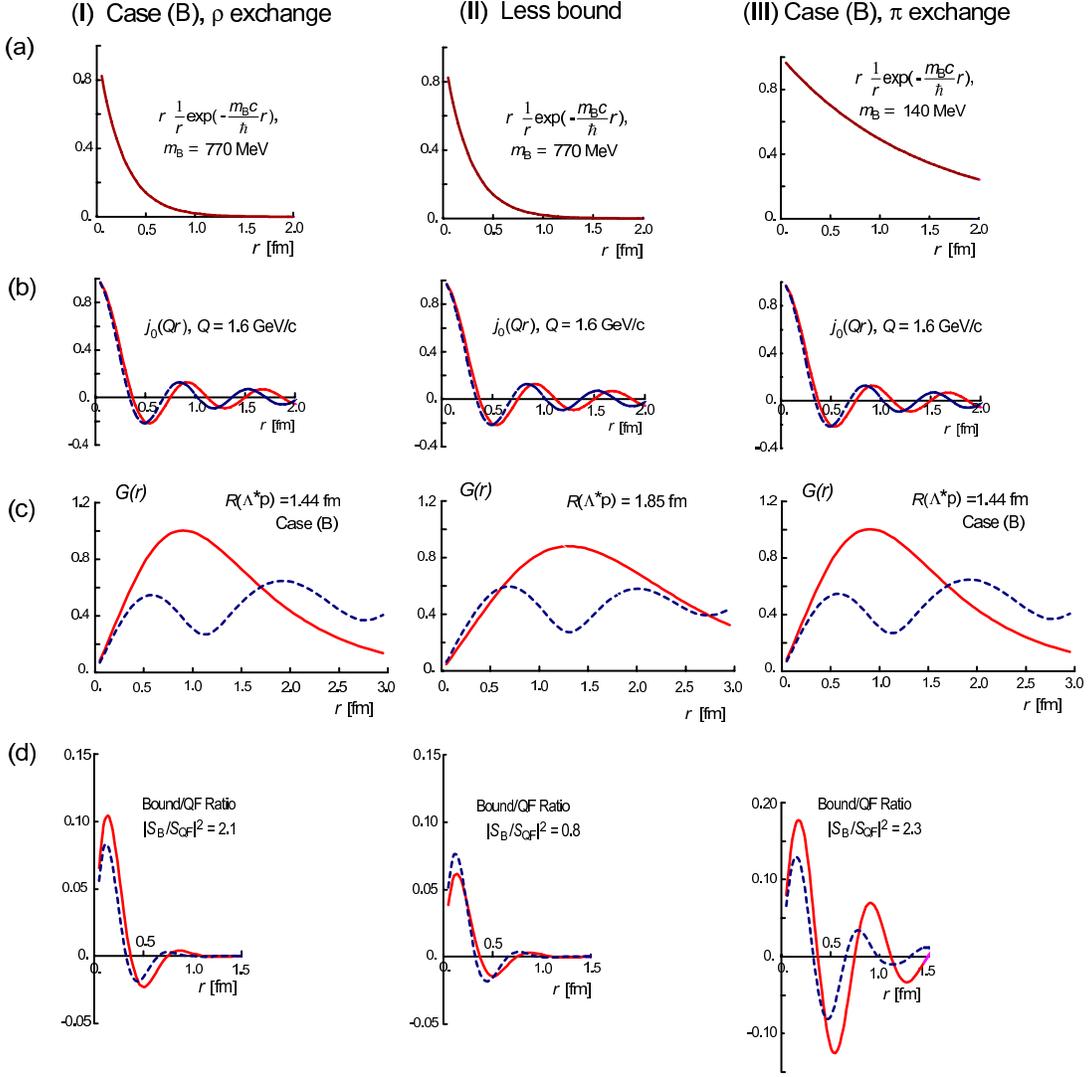}
\vspace{0cm}
\caption{\label{fig:T-amplitude} 
(Color online) Effects of various parameters on each factor of the transition amplitude in the $p + p \rightarrow K^+ + K^-pp$ reaction. Column (I) and (III) for the $\bar{K} N$ interaction Case (B) and with $\rho$ and $\pi$ exchange cases, respectively. Column (II) for the less bound case.  The solid (red) and broken (blue) curves represent the bound ($E_{\rm B}(\Lambda^*p) = -60$ MeV) and the quasi-free ($E_{\rm QF} (\Lambda^*p)$ = 100 MeV) regions. }
\end{figure*}

The production cross section of $\Lambda^*p$ ($= K^- pp$) at energy $E = (M-M_{\Lambda^*})c^2$ is given by
\begin{eqnarray}
&&\frac{{\rm d}^3 \sigma} {{\rm d} E {\rm d} \Omega_{K^+}} = \frac{(2\pi)^4}{(\hbar c)^2} |U_0|^2 \frac{k_{K^+} E_p}{2\, k_p} (-\frac{1}{\pi}) \times \nonumber\\
&&  {\rm Im} \big[\int \! \int {\rm d} \vec{r}' {\rm d} \vec{r} f^*(\vec{r}' ) \langle \vec{r}'|\frac{1}{E - H_{\Lambda^*p} + { i} \epsilon}|\vec{r} \rangle f(\vec{r}) \big]
\end{eqnarray}
where 
\begin{equation}
H_{\Lambda^* p} = -\frac{\hbar^2}{2 \, \mu_{\Lambda^* p}} \vec{\nabla}^2 + (v_0 + { i} w_0)\, r^2 {\rm exp} (-\frac{r^2}{c^2}) 
\end{equation}
is the $L^*$-$p$ interaction, represented with 
$v_0 = - 3770$ MeV, $w_0 = -880$ MeV, and $c = 0.3$ fm, which are deduced from the $N$-$(\bar{K}N)$ potential in Fig. \ref{fig:Structure-Kpp} obtained by the structure calculation of $K^-pp$.
 
Then, the spectral function of $\Lambda^*p$ ($= K^- pp$) is
\begin{eqnarray}
&&S(E) = - \frac{1}{\pi}  {\rm Im} \big[\int \! \int {\rm d} \vec{r}' {\rm d} \vec{r} f^*(\vec{r}') 
 \langle \vec{r}'|\frac{1}{E - H_{\Lambda^*p} + { i} \epsilon}|\vec{r}\rangle f(\vec{r}) \big] \nonumber\\
&& = -\frac{8 \mu_{\Lambda^* p}}{\hbar^2} \sum_{l=0}^{\infty} (2 l +1) {\rm Im}  \big[\frac{1}{W(u_l^{(0)} u_l^{(+)})} \nonumber\\
&& \times \int_{0}^{\infty} {\rm d}r' \int_{0}^{\infty} {\rm d}r\, {\rm exp}(-\frac{r'}{\beta}) \, j_l (Q r') \nonumber\\
&& \times~  u_l^{(0)} (r_<) u_l^{(+)} (r_>) \, j_l (Qr) ~{\exp}(-\frac{r}{\beta}) \big],
\end{eqnarray}
where $u_l^{(0)}$ and  $u_l^{(+)}$ are the stationary and outgoing solutions of the Schr\"odinger equation,
and $W$ is the Wronskian of them.

\begin{figure*}
\centering
\includegraphics[width=16cm]{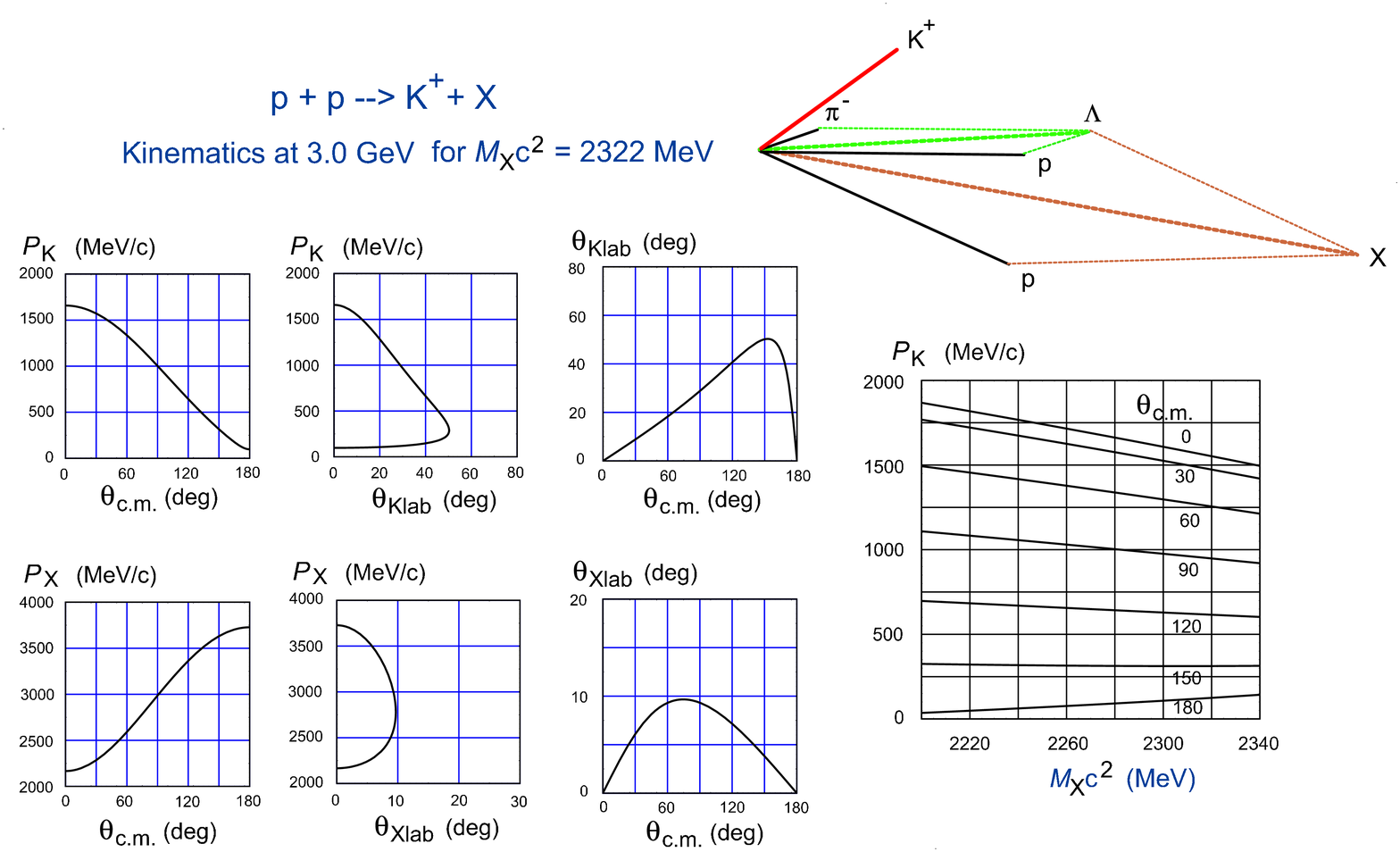}
\vspace{0cm}
\caption{\label{fig:pp2KX-kinematics} 
(Color online) Kinematic relations in $pp \rightarrow K^+ + X$ reaction at $T_p$ = 3.0 GeV for $M_X = 2280$ MeV/$c^2$. The momentum vectors for a typical backward $K^+$ event are also shown.
}
\end{figure*}

  Essentially, the spectral function is composed of the following three factors:
\begin{equation}
  \frac{{\rm e}^{-m_B\, r}}{r} \times {\rm e}^{i \vec{Q} \vec{r}} \times G (r), \label{eq:transition-intensity}
\end{equation}
where
\begin{equation}
G (r) = \big[ -{\rm Im} \{\frac{u^{(0)} (r)\, u^{(+)} (r)}{W(u^{(0)} \, u^{(+)} )}  \} \big]^{1/2}.
\end{equation}
They are: i) the collision range $1/m_B$, ii) the momentum transfer $Q$ and iii) the structure function $G(r)$ depending on the rms distance $R(\Lambda^* p)$ of the $\Lambda^*$-$p$ system. The calculated wavefunction of $K^- pp$ in the case (B) yields $R(\Lambda^* p) =$ 1.44 fm. The momentum transfer in the reaction is $Q \sim 1.6$ GeV/$c$, depending on the angle. The boson mass in producing $\Lambda^*$ in $pp$ collision is taken to be the $\rho$ meson mass; $m_B =  m_{\rho} =  770$ MeV/$c^2$. (For comparison we also examined the case of $m_B =  m_{\pi} =  140$ MeV/$c^2$, and found a similar result.) 

The calculated spectral function at $T_p$ = 4 GeV at forward angle in the scale of $E(\Lambda^* p) = 27~{\rm MeV} - B_K$ is presented in Fig.~\ref{fig:piK-pp-combined} (Lower-Right). Surprisingly, in great contrast to the ordinary reactions, the spectral function is peaked at the bound state with only a small quasi-free component. This means that the sticking of $\Lambda^*$ and $p$ is extraordinarily large.

\subsection{Unique features of the $p$-$p$ reaction: dominance of the $\Lambda^* p$ doorway}

This dominance of $\Lambda^* p$ sticking in such a large-$Q$ reaction can be understood as originating from the matching of the small impact parameter with the small size of the bound state. For further understanding of the mechanism we examined the dependence of the spectral function by changing the essential parameters fictitiously. Figure~\ref{fig:piK-pp-combined} (Lower-Right) shows that the bound-state peak decreases, when we increase the rms size $R(\Lambda^* p)$ from 1.44 fm (the predicted size in the case (B)) to 1.67 fm (case (A)), and further to 1.85 fm. It also shows that with a hypothetically denser system ($R(\Lambda^* p)$ = 1.25 fm) the peak height increases dramatically. So, we prove that the dominant sticking of $\Lambda^* p$ is the result of the dense $\bar{K}$ system to be formed. 

As a more analytical way to show the physics behind we plot in Fig.~\ref{fig:T-amplitude} the radial dependence of each factor of the transition intensity, eq.(\ref{eq:transition-intensity}), for different cases of the parameters, Column (I) for Case (B) and $\rho$ exchange, Column (II) for less bound case, and Column (III) for Case (B) and $\pi$ exchange. The first row (a) shows the $p$-$p$ interaction range of the Yukawa type with different boson masses $m_B$. The second row (b) shows the spherical Bessel function $j_0 (Qr)$ corresponding to the momentum transfer of $Q =$ 1.6 GeV/$c$. The third row (c)  shows the structure dependent function $G(r)$, where the solid (red) and broken (blue) curves represent the bound ($E_{\rm B}(\Lambda^*p) = -60$ MeV) and the quasi-free ($E_{\rm QF} (\Lambda^*p)$ = 100 MeV) regions, respectively. Finally, the bottom row (d) shows the radial dependences of the spectral strengths at the bound-state and the quasi-free (QF) regions. In the case (I), the structure function is dense for $r < 1.5$ fm so that it overlaps with the short range interaction (a) assisted by the large momentum transfer (b). The spectral intensity damps and shows little oscillation after $r \sim$ 0.5 fm. Thus, the radial-integrated spectral intensity is large, and the Bound/QF ratio is also large, $\sim$ 2.1. If we artificially increase the $\Lambda^*p$ distance to 1.85 fm, as shown in Column (II), the initial overlapping part drops down so that the ratio becomes 0.8. A softened $p-p$ interaction ($m_B =$ 140 MeV), Column (III), makes the spectral intensities more oscillatory, yielding smaller intensities, but the Bound/QF ratio (=2.3) is unchanged from the case (I).

We also checked that with a fictitiously long-range $NN$ collision ($m_B$ = 10 MeV) the bound-state peak diminishes and the quasi-free component dominates. Under this condition the bound-state peak is enhanced only when the momentum transfer is small (so called recoilless condition). On the other hand, the dominant sticking of $\Lambda^* p$ is assisted by the large momentum transfer ($Q \sim $ 1.6 GeV/$c$). 

We have thus demonstrated that the dominant sticking of $\Lambda^* p$ occurs as a joint effect of the short-range collision, the large momentum transfer and the compact size of the $\bar{K}$  cluster. It is vitally important to examine our results experimentally. An experimental observation of $K^- pp$ in $pp$ collision will not only confirm the existence of $K^- pp$, but also proves the compactness of the $\bar{K}$ cluster.

\subsection{Production cross sections and kinematics}

The reaction we propose is essentially a reaction of two-body final states,  
\begin{equation}
p + p \rightarrow K^+ + X,
\end{equation}
where the unknown object $X$ with a mass $M_X$, can be searched for in a missing mass spectrum of $K^+$, $MM(K^+)$. For a given $M_X$ the $K^+$ momentum in the laboratory frame is a unique function of the c.m. angle. With an incident energy of $T_p = 3.0$ GeV we calculated kinematical relations of the above reaction, as shown in Fig.~\ref{fig:pp2KX-kinematics}. The laboratory angle of $K^+$ spans from 0 to a maximum (around 50 degrees,), where the c.m. angle is 75 degrees. The momentum transfer $Q$ is around 1.6 GeV/c.

Parallel to the missing-mass spectroscopy we can perform invariant-mass spectroscopy of $(\Lambda p)$ pair as well. Figure~\ref{fig:pp2KX-kinematics} also shows a typical event pattern, where a $K^+$ is emitted at large angle with a moderate momentum (500 - 1000 MeV/c). The corresponding $X$ goes out at a forward angle (both in c.m. and in lab.) and thus its decay particles, $\Lambda + p \rightarrow p + \pi^- + p$, are forward boosted. 

\begin{figure}
\centering
\includegraphics[width=\columnwidth]{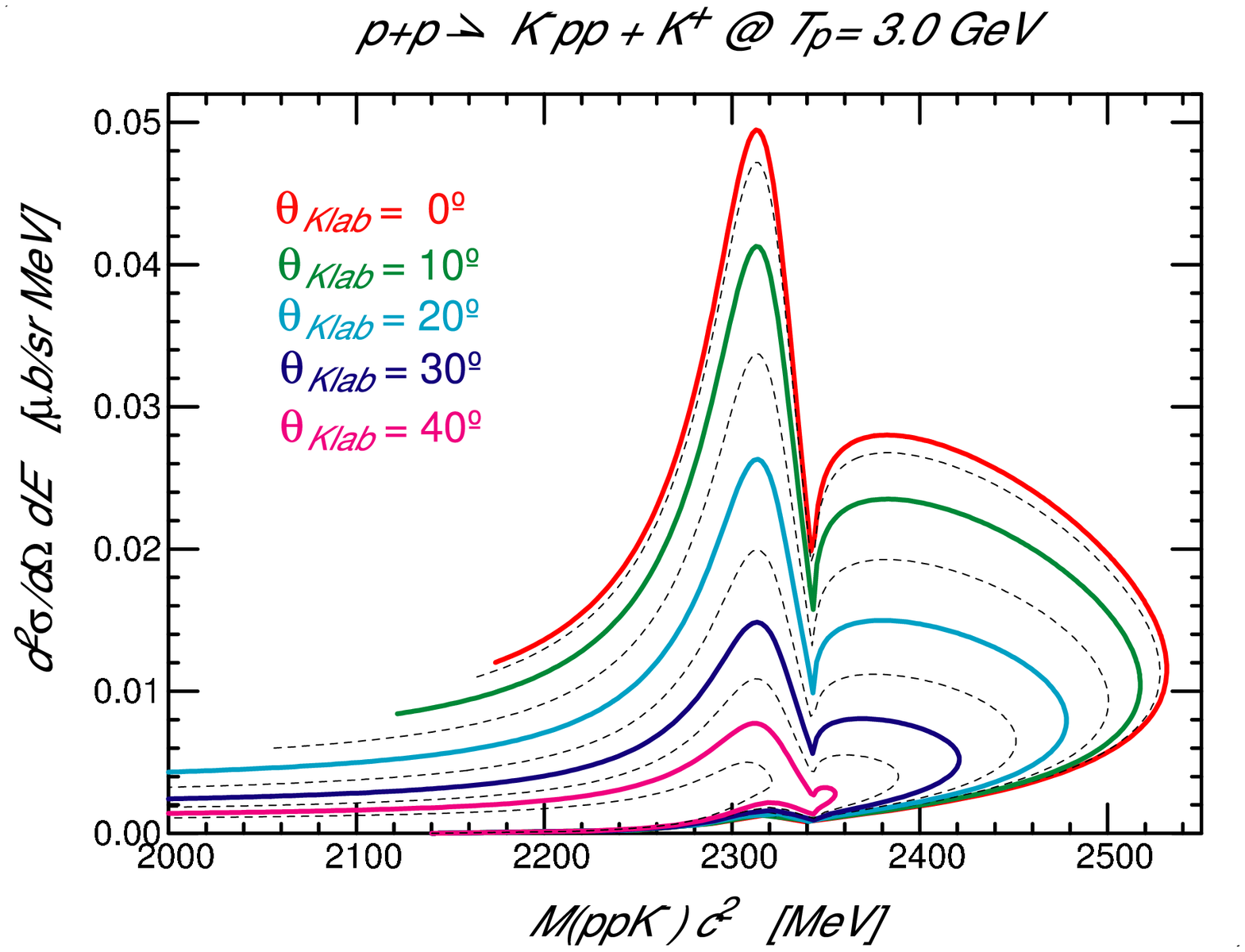}
\vspace{0.5cm}
\includegraphics[width=\columnwidth]{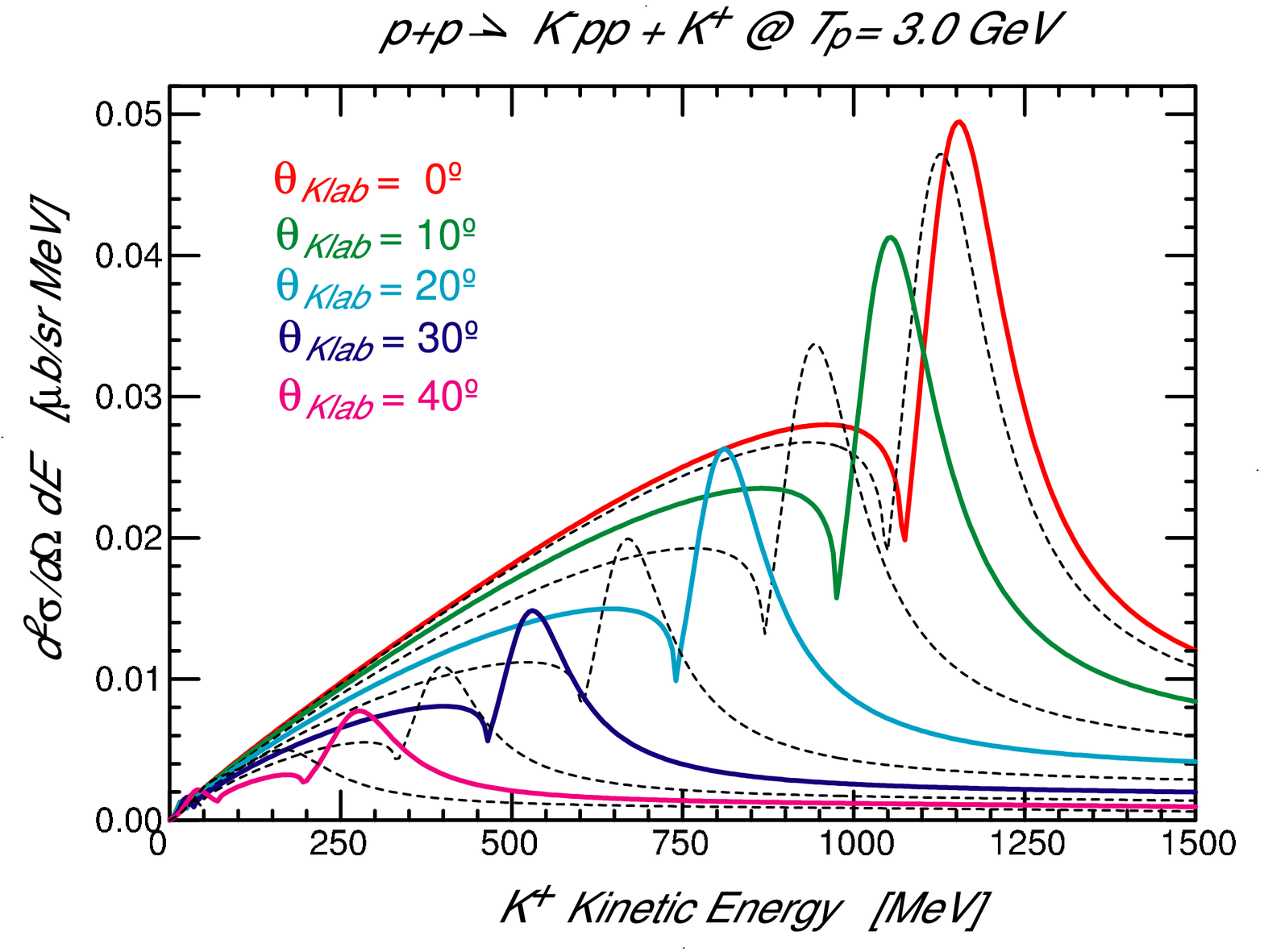}
\vspace{0cm}
\caption{\label{fig:Cross-section} 
(Color online) Predicted differential cross sections of $p + p \rightarrow K^+ + X$ at $T_p = 3.0$ GeV for the $\bar{K}N$ interaction Case (A). (Upper) $M_X$ spectra at various $K^+$ laboratory angles. (Lower) $K^+$ energy spectra at various $K^+$ laboratory angles. }
\end{figure}

The calculated cross sections at $T_p$ = 3 GeV at various laboratory angles for an assumed bound-state mass of $M_{K^-pp}$ = 2310 MeV/$c^2$, the original case (A), are presented in Fig.~\ref{fig:Cross-section}. The upper one is in the scale of $M_X$. The cross section integrated over the quasi-free region ($E(\Lambda^* p) > 0$) corresponds to the free emission of $\Lambda^*$ above 2340 MeV/$c^2$, which is known to have an empirical cross section of  $\sigma (pp \rightarrow K^+ + \Lambda^* + p) =$ 20 $\mu$b from a DISTO experiment at $T_p$ = 2.85 GeV \cite{DISTO}.  So, we have adjusted our absolute cross sections so as to give this empirical cross section.

The cross section has substantial angular dependence, but the bound-state peak is distinct at any angle. Even at large laboratory angle around 30 degrees the cross section is modest and the peak to background ratio remains large.  The cross section in the scale of $K^+$ energy is shown in the lower part of Fig.~\ref{fig:Cross-section}. 

\begin{figure}
\centering
\includegraphics[width=\columnwidth]{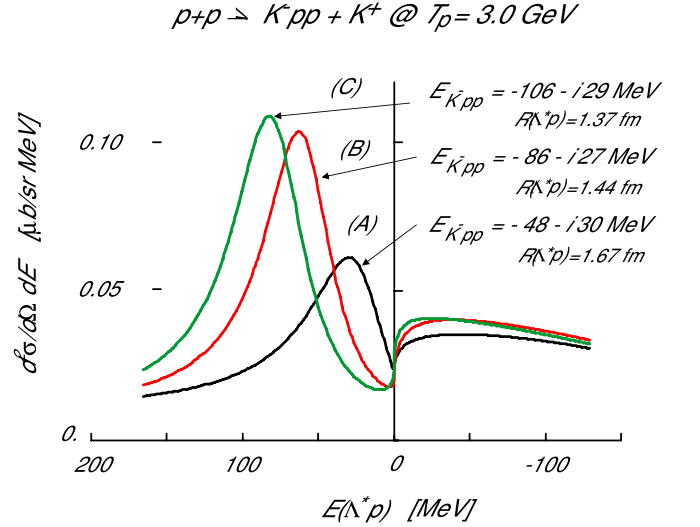}
\vspace{0cm}
\caption{\label{fig:Cross-section-ABC} 
(Color online) Predicted differential cross sections of $p + p \rightarrow K^+ + X$ at $T_p = 3.0$ GeV at forward angle  in the three cases of the $\bar{K}N$ interaction: Case (A), (B) and (C), as given in Table~\ref{tab:KNN}. The $E(\Lambda^* p) = 0$ value corresponds to the $\Lambda^*$ emission threshold.}
\end{figure}

 Figure~\ref{fig:Cross-section-ABC} shows the cases of the original (A) and enhanced interactions, (B) and (C), as presented in Table~\ref{tab:KNN}. The peak position moves toward lower masses and the peak cross section increases accordingly.

The elementary reaction of type 
\begin{equation}
p + p \rightarrow K^+ + Y^0 + p
\end{equation}
was studied experimentally by the DISTO group at SATURNE \cite{DISTO} and more recently by the ANKE group at COSY \cite{Zychor:06}. The DISTO experiment identified $\Lambda$ from  the invariant-mass spectrum of $p + \pi^-$ and constructed a missing mass spectrum of $K^+ p$ from those events involving $\Lambda$ at an incident proton energy of $T_p$ = 2.85 GeV. The missing mass in this case corresponds to the mass of $Y^0$, 
\begin{equation}
MM(K^+p) = M(Y^0),
\end{equation}
and they found peaks associated with the production of $\Lambda(1115)$, $\Sigma^0 (1193)$ and ${\Sigma^0} (1385) + \Lambda (1405)$. They obtained a cross section of 20 $\mu$b for $\Lambda (1405)$, which we have used as an absolute scale in our calculations.

The ANKE experiment measured the energies and momenta of four emitted particles, $K^+$, $p$, $X^{+,-}$ and $\pi^{-,+}$, with complete kinematical constraint at $T_p$ = 2.83 GeV:
\begin{equation}
p + p \rightarrow K^+ + p + Y^0 \rightarrow K^+ + p + X^{+,-} + \pi^{-,+}.
\end{equation}
They constructed $MM(K^+ p)$ and additionally $MM(K^+ p \pi^-)$, which was equated to $M(X)$. When $MM(K^+ p \pi^-) = M(p)$, it indicates that this $Y^0$ decays to $p + \pi^-$, and thus, it is assigned to $\Lambda$. Thus, the ANKE $MM(K^+ p)$ spectrum is similar to the DISTO spectrum. 

In these experiments the reaction products were studied in terms of the elementary processes. Now, we propose to examine the new situation related to the existence of $K^- pp$. We point out that the formation/decay process of this object, as given in eq.(\ref{eq:pp2KLp}) and (\ref{eq:pp2Kpp-decay-1},\ref{eq:pp2Kpp-decay-2},\ref{eq:pp2Kpp-decay-3}), are hidden in the observed spectra of $MM(K^+ p)$. The most important information in our context is contained in a spectrum of $MM(K^+)$, which is related to the mass of $K^- pp$,
\begin{equation}
MM(K^+) = M(K^- pp),
\end{equation}      
but no such spectrum has been reconstructed yet. 
Now, a new experiment of the FOPI group at GSI \cite{FOPI-proposal}, which is aimed at measuring the whole products in the $p + p$ reaction at $T_p$ = 3 GeV to reconstruct both the invariant mass $M_{\rm inv} (\Lambda p)$ and the missing mass $MM(K^+)$, is in progress.

\subsection{Subsequent $\Lambda^*p$ doorway processes}

Once a $\Lambda^* p$ doorway is formed, it proceeds to a bound $K^-pp$ state, and thus it is likely to further propagate in a complex nucleus as
\begin{eqnarray}
&&\Lambda^* p + ``p" \rightarrow K^-ppp,\\
&&\Lambda^* p + ``n" \rightarrow K^-ppn,
\end{eqnarray}
where the $K^-$ traverses through the three nucleons coherently, and ultimately a {\it kaonic proton capture} reaction may occur, such as
\begin{eqnarray}
&&d(p,K^+)K^-ppn,\\
&&d(p,K^0)K^-ppp,\\
&&^3 {\rm He}(p,K^+)K^-pppn,\\
&&^3 {\rm He}(p,K^0)K^-pppp.
\end{eqnarray}
In principle, missing mass spectra, $MM(K^+)$ and $MM(K^0)$, may reveal monoenergetic peaks.

\section{Concluding remarks}

In this paper we have presented the results of our comprehensive three-body calculations on the structure of the basic $\bar{K}$ cluster, $K^-pp$. First, we have shown that the single-channel complex potential of $\bar{K}N$, transformed from coupled-channel interactions, has very little energy dependence, which justifies our three-body calculations. The binding energy and width of $K^-pp$, which were obtained in our original prediction, are found to be robust against any change of the $\bar{K}N$ and $NN$ interaction parameters, as far as they account for $\Lambda(1405)$ as a $K^-p$ bound state. It is also shown that the structure of $K^-pp$ can be interpreted as a covalent state of $p$-$K^-$-$p$, in which the $K^-$ migrates over the two protons coherently, yielding ``super strong" nuclear force. In this sense, the $K^-pp$ cluster may be called a ``mini sub-femtometer dense diatomic molecule" with a quasi $\Lambda(1405)$ as the ``atom". This structure justifies our $\Lambda^*$ doorway treatment of the formation reaction. The $K^-pp$ as a dissolved state of $\Lambda^* + p$ is predicted to be formed with extraordinarily large enhancement, since the $\Lambda^*$ produced in a short-range collision with the participating proton spontaneously forms a $\Lambda^* p$ doorway, which is nearly equivalent to the dense $K^-pp$ system. This anomalous dominance of $\Lambda^* p$ sticking is shown to result from the unusual matching of the short  collision range ($1/m_B \sim$ 0.3 fm) and the small radius of the produced $K^- pp$, assisted by a large momentum transfer. Experimental confirmation of this effect will simultaneously prove the compact character of the $\bar{K}$ cluster. 

Finally, we comment on our earlier proposal to make use of hot fireball in heavy-ion collisions as sources of various $\bar{K}$ clusters \cite{Yamazaki:04}. If the formation of $\Lambda^* p$ doorway toward $K^-pp$ is really enhanced in $NN$ collisions, the successive formation of $\bar{K}$ clusters, $K^- pp$, $K^- ppp$, etc., in heavy-ion reactions is expected to be enhanced as well, since a heavy-ion reaction involves lots of primary $NN$ collisions and subsequent $\Lambda^* N$ production processes. \\ 

We would like to thank Prof. P. Kienle, Prof. N. Herrmann, Dr. K. Suzuki and Dr. L. Fabietti for the daily stimulating discussion during the course of the experimental collaboration, and Prof. S. Shinmura for the illuminating discussion. One of us (T.Y.) is grateful to the Alexander von Humboldt Foundation for its ``Forschungspreis".  We acknowledge the receipt of Grant-in-Aid for Scientific Research of Monbu-Kagakusho of Japan.

\clearpage\newpage

\end{document}